\documentclass{article}

\usepackage{PRIMEarxiv}

\usepackage[utf8]{inputenc} 
\usepackage[T1]{fontenc}    
\usepackage{hyperref}       
\usepackage{xurl}            
\usepackage{booktabs}       
\usepackage{amsfonts}       
\usepackage{nicefrac}       
\usepackage{microtype}      
\usepackage{lipsum}
\usepackage{fancyhdr}       
\usepackage{graphicx}       

\usepackage{float}
\graphicspath{{media/}}     

\usepackage{enumitem}

\usepackage{subcaption}

\usepackage{amsmath}

\usepackage[most]{tcolorbox}

\usepackage{comment}

\usepackage[disable]{todonotes}

\usepackage{listings}

\usepackage{tabularx} 
\usepackage{threeparttable}

\usepackage{pifont} 
\usepackage{xcolor} 

\newcommand{\cmark}{\textcolor{green}{\ding{51}}} 
\newcommand{\xmark}{\textcolor{red}{\ding{55}}}   
\newcommand{\omark}{\textcolor{yellow}{\ding{109}}}   

\title{Rendering Large Volume Datasets in Unreal Engine 5: A Survey
}

\author{
  Markus Schlüter, Tom Kwasnitschka, Armin Berstetter, Jens Karstens \\
  GEOMAR Helmholtz Centre for Ocean Research  \\
  Kiel\\
  \texttt{\{maschlueter, tkwasnitschka, abernstetter, jkarstens\}@geomar.de} \\
}

\begin{document}
\maketitle


\begin{abstract}

  In this technical report, we discuss several approaches to in-core rendering of large volumetric datasets in Unreal Engine 5 (UE5). We explore the following methods: the TBRayMarcher Plugin \cite{bib:TBRaymarcherPlugin2024}, the \href{https://dev.epicgames.com/community/learning/paths/mZ/unreal-engine-niagara-fluids}{Niagara Fluids Plugin}
  , and various approaches using
  \href{https://dev.epicgames.com/documentation/en-us/unreal-engine/sparse-volume-textures-in-unreal-engine}{\textit{Sparse Volume Textures (SVT)}}
  , with a particular focus on
  \href{https://dev.epicgames.com/documentation/en-us/unreal-engine/heterogeneous-volumes-in-unreal-engine}{\textit{Heterogeneous Volumes}} (HV)
  .
  We found the HV approach to be the most promising.

  The biggest challenge we encountered with other approaches was the need to chunk datasets so that each fits into volume textures smaller than one gigavoxel. While this enables display of the entire dataset at reasonable frame rates, it introduces noticeable artifacts at chunk borders due to incorrect lighting, as each chunk lacks information about its neighbors.

  After addressing some (signed) \lstinline{int32} overflows in the Engine's SVT-related source code by converting them to to (unsigned) \lstinline{uint32} or \lstinline{int64}, the SVT-based HV system allows us to render sparse datasets up to $32k \times 32k \times 16k$ voxels, provided the
  compressed tile data (including MIP data and padding for correct interpolation )
  does not exceed 4 gigavoxels.

  In the future, we intend to extend the existing SVT streaming functionality to support out-of-core rendering, in order to eventually overcome VRAM limitations, graphics API constraints, and the performance issues associated with 64-bit arithmetic in GPU shaders.

\end{abstract}


\begin{figure}[H]
  \centering
  \includegraphics[width=1.0\textwidth]{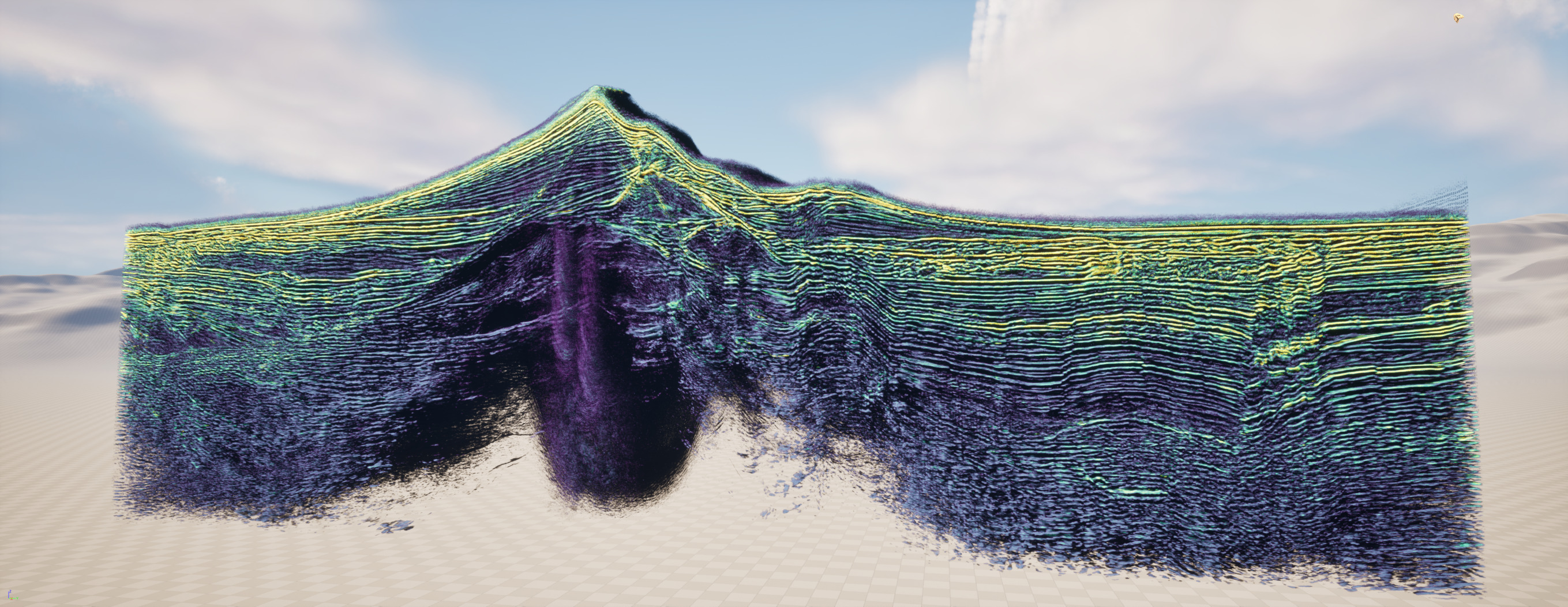}
  \caption{The Kolumbo dataset rendered as a Heterogeneous Volume in a custom build of Unreal Engine 5.4,
    achieving a frame rate of 20 fps on an NVIDIA RTX 3500 Ada Generation Laptop GPU. The total (dense) volume resolution is
    $4211 \times 1501 \times 935 \approx 6 $  billion voxels, while it has about 2.6 billion non-empty voxels.}
  \label{fig:Kolumbo_HetVol_full_fixed}
\end{figure}




\section{Introduction}

Video game engines have matured significantly in terms of feature richness, visual fidelity and performance, and are increasingly adopted for applications beyond game development \cite{bib:UnrealBeyondEntertainment}.

\begin{figure}[H]
  \centering
  \includegraphics[width=1.0\textwidth]{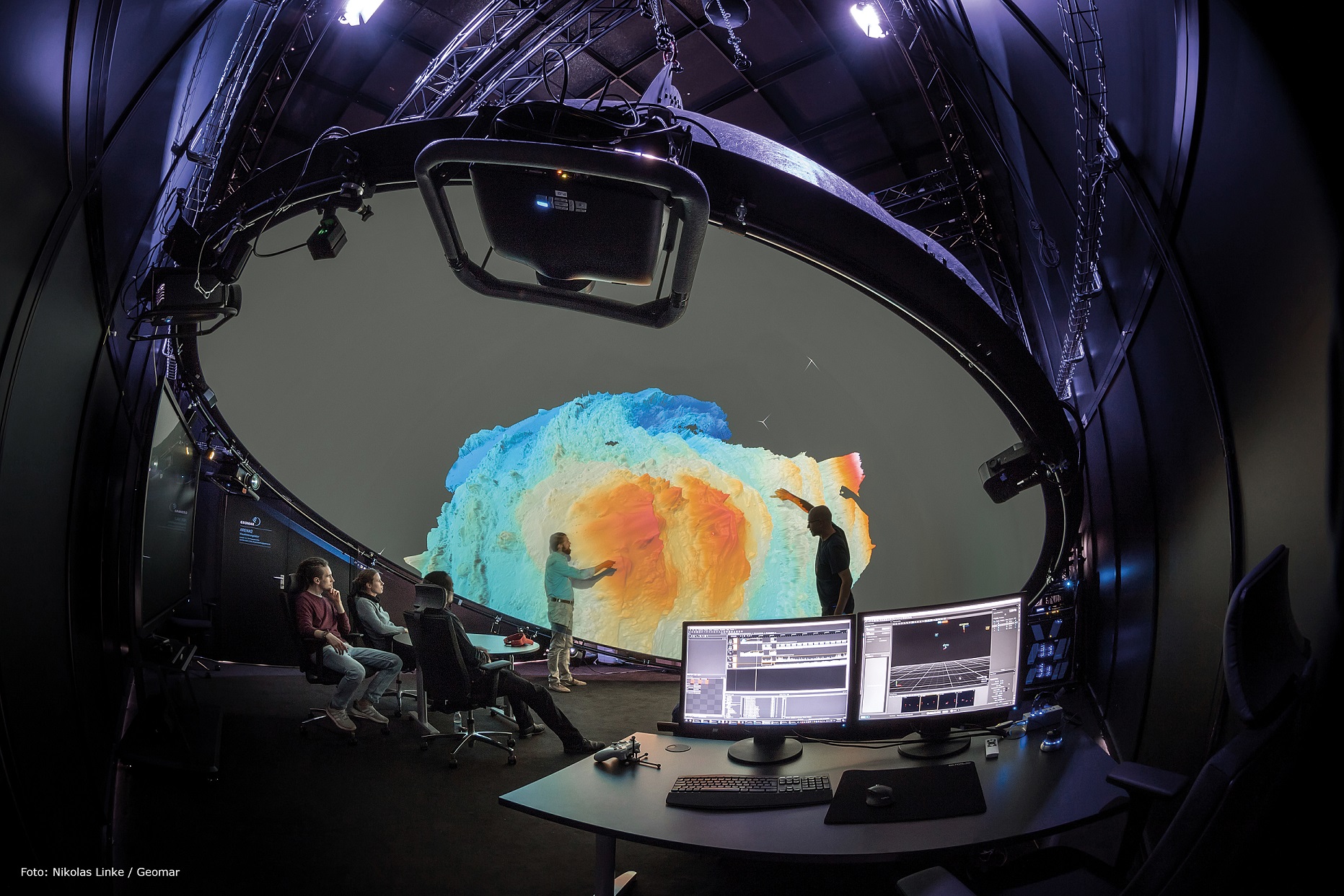}
  \caption{The ARENA2 visualization dome at the GEOMAR Helmholtz Center for Ocean Research Kiel}
  \label{fig:ARENA2_photo}
\end{figure}

Kwasnitschka et al. \cite{bib:kwasnitschka2023spatially} present our ARENA2 visualization dome (see fig. \ref{fig:ARENA2_photo}) at the GEOMAR Helmholtz Center for Ocean Research Kiel. Our aim is to provide an immersive, collaborative virtual environment for domain scientists to conduct, discuss and document their research.

In a distributed system like this, achieving performance and visual fidelity comparable to a desktop setup requires software capable of operating in cluster mode. This enables a distributed, synchronized rendering of a scene from multiple viewpoints. Each node renders an image from a specific pose and projects it—using warping and blending—to create a seamless image across the dome's hemisphere.

We have integrated several scientific visualization tools that support cluster mode, including ParaView, OpenSpace, CosmoScout VR, and the Digital Earth Viewer.
Additionally, more3D offers an approach that allows software without native cluster mode to render in multi-display setups by capturing, modifying, and distributing the buffers of the OpenGL graphics API. However, interactivity with this solution is limited.
\todo{cite any/each of them or provide URLs?}

Unreal Engine 5, with its \textit{nDisplay} technology \cite{ndisplay_whitepaper}, has established itself as a robust platform for multi-display immersive environments involving real-time computer graphics, including scientific visualization \cite{bib:UnrealBeyondEntertainment}. The engine is widely adopted and valued for its versatility, high visual fidelity, strong performance, active development, and extensive community support.

In the context of the
\href{https://bluehealthtech.de/}{Blue Health Tech (BHT)} project  \footnote{\url{https://www.bluehealthtech.de/}}
\href{https://www.hyperquant.de/}{Hyperquant}
\footnote{\url{https://www.hyperquant.de/}}
\todo{@Tom: cite BHT, show medical datasets?}
, a project aiming to combine expertise from ocean research with medical innovations, we require a solution to explore the volumetric medical data in our dome. As GEOMAR is an ocean research center, our aim is to generate synergies by providing an interactive, direct volume visualization solution that goes beyond the typical sizes of medical data sets. This would enable Unreal Engine’s features to be used by scientific disciplines that handle much larger volumetric datasets, e.g. seismology, geophysics in general and oceanography.

We have previously used Unreal Engine for other projects: For instance, Bernstetter \cite{bib:bernstetter2024virtualfieldworkimmersiveenvironments} provides tools for geologists to explore and measure geo-referenced bathymetry and LIDAR data. The results can be documented and reproduced using the \emph{Digital Lab Book}, a provenance tool integrated with Unreal Engine \cite{bib:bernstetter2023practical}. Adding support for large volumetric datasets is a natural extension of these capabilities.

This work aims to give an overview of the current volume rendering capabilities and limits of Unreal Engine 5.
As a side effect, it may provide some insights that could help with a future extension to true out-of-core rendering. \todo{do provide these insights by getting inspired by sciview.}

This report is structured as follows. First, we outline the requirements that a visualization solution should ideally meet in our scenario (Section \ref{sec:requirements}). Section \ref{sec:relatedwork} presents some background and related work in scientific visualization for multiprojection setups, such as CAVEs and domes, particularly for volumetric datasets. In Section \ref{sec:approaches}, we discuss the approaches we evaluated: creating a custom solution, using plugins, adapting existing systems, and exploring an experimental feature. We present and discuss our results in Section \ref{sec:results}, conclude with an outlook in Section \ref{sec:conclusion}.

\begin{tcolorbox}[
    colback=blue!2!white,
    colframe=white!85!black,
    title=Notation,
    fonttitle=\color{black} 
  ]
  As we deal with sizes and limits that are best expressed as powers of two,
  when using scales like kilo-, mega- or giga- voxels of bytes, we refer to the \textbf{power-of-two} scale (kilo$=2^{10}$, mega$=2^{20}$, giga$=2^{30}$),
  in contrast to the power-of-ten scale (kilo$=10^{3}$, mega$=10^{6}$, giga$=10^{9}$).
  When we express a number with the power-of-ten scale, we either make it explicit via
  scientific notation (e.g. $a \times 10^{x}$) or by naming them (thousand, million, billion).
  This is just a technical remark; for a general understanding, the order of magnitude often is more important than the exact value, anyway.
\end{tcolorbox}


\section{Requirements}
\label{sec:requirements}

We are looking for an interactive, accurate, visually pleasing, and extensible direct visualization solution of large voxel-based volumetric data in a multiprojection dome.

Krüger et al. \cite{bib:UnrealBeyondEntertainment} specify "six fundamental requirements that software solutions should satisfy to serve as a robust foundation for development":
\begin{enumerate}
  \item Wide Adoption \label{req:Adoption}
  \item Large Feature Set \label{req:Feature}
  \item Performance  \label{req:Performance}
  \item Accessibility \label{req:Accessibility}
  \item Extensibility/Adaptability \label{req:Extensibility}
  \item Flexibility \label{req:Flexibility}
\end{enumerate}

The authors conclude that the Unreal Engine fully fulfills the req.  \ref{req:Adoption}, \ref{req:Extensibility}  and \ref{req:Flexibility} , mostly meets req. \ref{req:Feature}, and meets the requirements \ref{req:Performance}  and \ref{req:Accessibility} under certain conditions. They conclude to continue using UE in their CAVE research environment and mention their interest in deeper knowledge of certain engine features to better make informed decisions.

Depending on the technical depth, there exist varying degrees of documentation of the engine: On the one hand, there is a vast amount of resources on how to use the engine from a game developer's perspective, but the deeper you want to dig into the internals of the C++ source code of the engine itself, documentation gets more and more sparse or less accessible.

So, despite some documentation, the true potential and limits of some approaches were not obvious from the beginning.

This report shall shine some light on the engine's volume rendering capabilities and - to a certain extent - its technical background. This may assist in getting an idea on how new features may be implemented in the future, particularly out-of-core volume rendering.

In our experience, the requirements \ref{req:Accessibility} and \ref{req:Extensibility}, accessibility and extensibility/adaptability, which according to  \cite{bib:UnrealBeyondEntertainment}  are met under conditions, also have turned out to be a double-edged sword: While some features worked right out of the box or could be easily modified, some others took a deep analysis of the engine's source code to get an idea of what is going on and how to possibly extend and/or modify it.

In the particular use case of volumetric rendering, we have identified the following specific requirements:

\begin{enumerate}[resume]
  \item Dataset size limits, if any
        \label{req:size}
  \item Display Accuracy: Compression shall be lossless
        \label{req:Accuracy}
  \item Compatibility to and Synergy with related projects
        \label{req:Compatibility}
  \item Explorability: Transfer function Specification, Interaction with the Data
        \label{req:Explorability}
  \item Sustainability: A solution should have minimal maintenance overhead
        \label{req:Sustainability}
\end{enumerate}

Ideally, we would like to render in real time (req. \ref{req:Performance}) data of arbitrary size (req. \ref{req:size}) in a visually expressive and appealing way (req. \ref{req:Feature}) without any bias, i.e. high accuracy (req. \ref{req:Accuracy}) : no loss of precision due to lossy data compression, subsampling, missing values, or lower numerical precision than the original data.
In practice, there are limitations in terms of storage, RAM and VRAM sizes, processing speed, memory bandwidth, API and hardware limits, requiring some compromises depending on the type of data and the use case.

First, we encountered a limit for allocating large volume textures at about 1 gigavoxel,
probably due to memory fragmentation issues with DirectX 11, which requires \emph{contiguous} memory in VRAM  for each resource.
This prohibits using larger volume textures even if there is enough free VRAM to fit a much larger volume texture, because the VRAM may be fragmented.
On the other hand, DirectX 12 allows for creating textures of up to $2048^3=8G$ voxels not only as a theoretical API limit,
but using \href{https://learn.microsoft.com/en-us/windows/win32/direct3d12/memory-management-strategies}{\textit{reserved resources}}
\footnote{\url{https://learn.microsoft.com/en-us/windows/win32/direct3d12/memory-management-strategies}}
,
the backing memory can actually be allocated as a collection of many tiles.
The virtual address space of the large volume texture is then mapped to these allocated tiles, allowing a huge texture allocation even with fragmented VRAM.
Knowing that UE supports DirectX 12 reserved resources, we want to be able to make use of volume textures that can be as large as 8 gigavoxels as much as possible.

Furthermore, we would like to profit from synergies with our other projects like \cite{bib:bernstetter2024virtualfieldworkimmersiveenvironments} and \cite{bib:bernstetter2023practical}, e.g. to overlay different datasets or use the provenance tool (req. \ref{req:Compatibility}).

Depending on the scientific domain and data types, a user may want to interact with the data in a specific way, like selecting and displaying numerical values or highlighting certain features (req.s \ref{req:Explorability}, \ref{req:Feature}, \ref{req:Extensibility} and \ref{req:Flexibility}).

Req. \ref{req:Accuracy} discourages the use of otherwise interesting approaches such as lossily compressing and reconstructing volume data, e.g. using neural networks proposed by Kim et al. \cite{bib:kim2024neuralvdb}.

Last but not least, a solution shall be easy to maintain and extend, particularly with regard to updates of its dependencies (req. \ref{req:Sustainability}).


\section{Background and Related Work}
\label{sec:relatedwork}

In this section, we discuss related work in the field of scientific visualization, focusing on multi-projection environments such as CAVEs and domes and volumetric datasets, and deliver some context on in-house development, datasets and hardware.

Kwasnitschka et al. \cite{bib:kwasnitschka2023spatially} describe our ARENA2 visualization dome, its purpose and projects, where the most recent ones are based on the Unreal Engine.

As mentioned in the introduction, Bernstetter  \cite{bib:bernstetter2024virtualfieldworkimmersiveenvironments} describes a tool
to explore and measure bathymetry and 3D mesh data reconstructed from LIDAR data in the ARENA2 lab using Unreal Engine and its Cesium plugin \todo{reference cesium?}. The data is represented as georeferenced meshes or as Cesium tiles.
The data sets involve, among others, bathymetric data of the Kolumbo submarine volcano near Santorini, Greece. As the corresponding survey also includes 3D seismic data, adding support for large volumetric datasets to the mix is a natural extension, and the one created by Karstens et al. \cite{bib:karstens2023cascading} serves as an exemplary dataset for this work. It is a 24GB SEG-Y file and has dimensions of $4211 \times 935 \times 1501$, that is, 5.5 billion voxels with a scalar 32bit floating point value per voxel, which encodes a seismic amplitude \todo{Ask Jens about physical units?}.

Thrastarson et al. \cite{bib:REVEAL_10.1785/0120230273} present REVEAL, a global-scale seismic model, which uses worldwide seismic measurements of earth quakes to derive properties such as wave propagation velocity. We used this 3D data set to visualize it directly, in contrast to 2D plotting of depth layers. \todo{Omit the mention of REVEAL?}

Krüger et al. \cite{bib:UnrealBeyondEntertainment} report their experiences with Unreal Engine in a scientific visualization CAVE. They argue mainly in favor of using the engine in this context, but identify some areas for improvement (see Section \ref{sec:requirements}).

Chang \cite{bib:Chang2020} implements direct volume rendering in Unreal Engine 4 in a medical context.
We got access to the source code that is not publicly available, and managed to get the project up and running, but we did not evaluate the solution any further, because of requirement \ref{req:Sustainability}: sustainability: Although performance, feature set, and visual fidelity look promising, the project would have needed to be migrated to Unreal Engine 5.

Bazar's \textit{TBRayMarcher} project\cite{bib:TBRaymarcherPlugin2024} also started by implementing direct volume rendering of medical datasets in Unreal Engine 4. It is developed as a plugin to the engine, with a stronger emphasis on virtual reality features.
Like \cite{bib:Chang2020}, it also delivers high frame rates, pleasant visuals, and a selection of transfer functions. Additionally, thanks to the explicit VR support, it has a GUI that we were able to use in our dome with some modifications. The project has also been updated for each engine version since UE 4.26, up until the most recent version 5.4. These features make the project a strong candidate for further evaluation; see section \ref{subsec:tbrm}.

\textit{NVIDIA IndeX} is a commercial software solution that leverages GPU clusters for real-time scalable visualization and computing of multi-valued volumetric data together with embedded geometry data \cite{nvidia_index_whitepaper}.
There is also a plugin for \textit{ParaView} \cite{nvidia_index_paraview}. This approach sounds quite promising, as IndeX allows for out-of-core rendering of large volume data sets, and ParaView enables the multichannel rendering required for the dome.

\textit{scenery/sciview} \cite{bib:scenery/sciview} is a scientific visualization framework written in Kotlin running on a Java VM. According to the paper, it features CAVE support and out-of-core volume rendering. Apart from not being based on Unreal Engine, this makes the project an interesting option.

\subsection{Hardware Setup}

The five render nodes of our visualization dome each have 32 GB RAM and an NVIDIA RTX 5000 GPU with 16 GB VRAM, which are synchronized via NVIDIA QuadroSync cards.
The development machine is a Dell Precision 7680 laptop with 32 GB RAM and an NVIDIA RTX 3500 Ada Generation.


\section{The Different Approaches}
\label{sec:approaches}

In this section, we present the approaches we explored to bring interactive, visually expressive and - where possible - accurate direct volume rendering of large voxel-based data into the ARENA2 dome.

\subsection{Custom Implementation}
\label{subsec:ownimpl}

\begin{figure}[H]
  \centering
  \includegraphics[width=0.5\textwidth, angle=180]{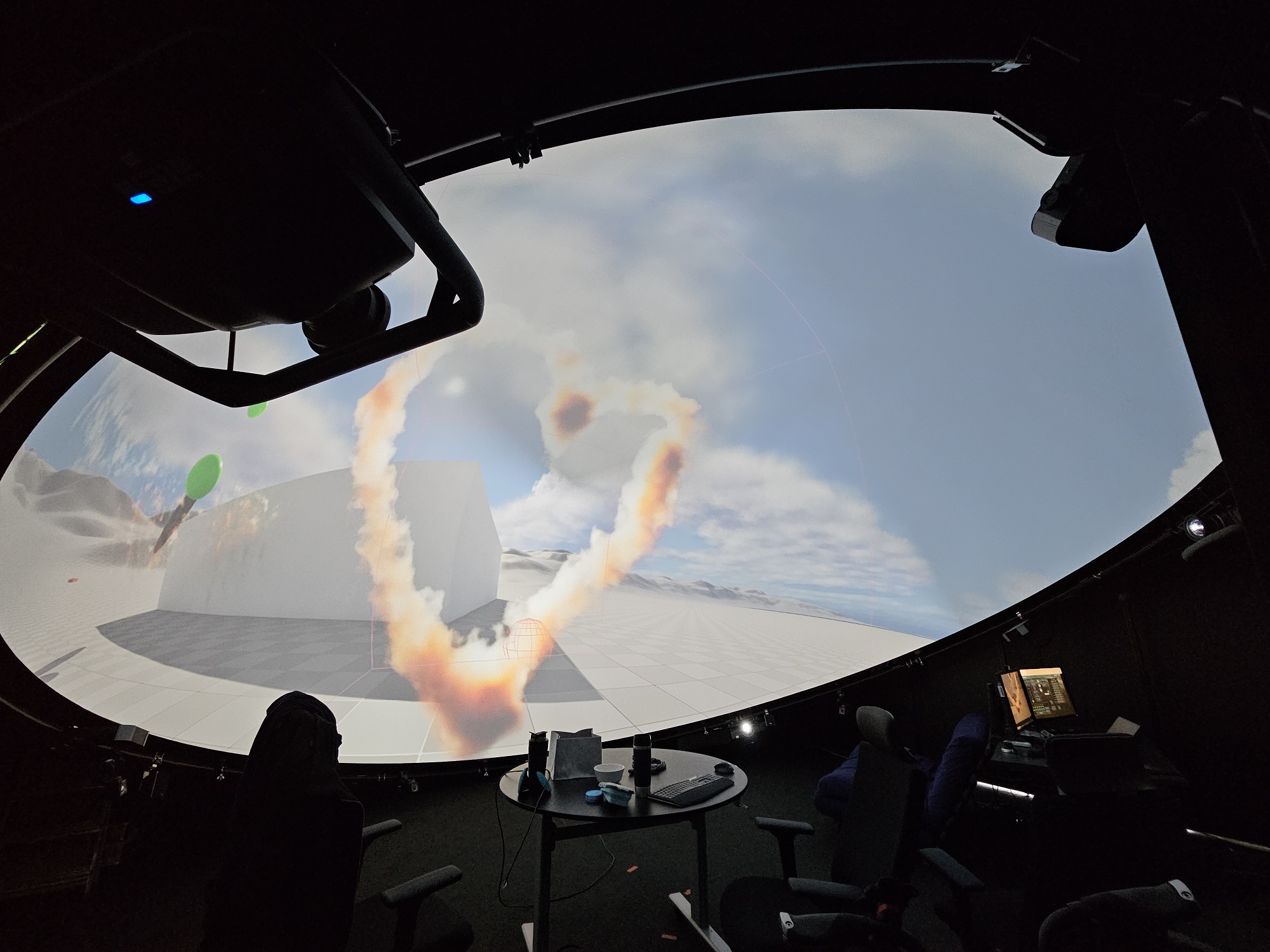}
  \caption{"Painting" a volume in our ARENA2 visualization dome.}
  \label{fig:VolumeRenderingFromScratch}
\end{figure}

\begin{figure}[H]
  \centering
  \includegraphics[width=0.75\textwidth]{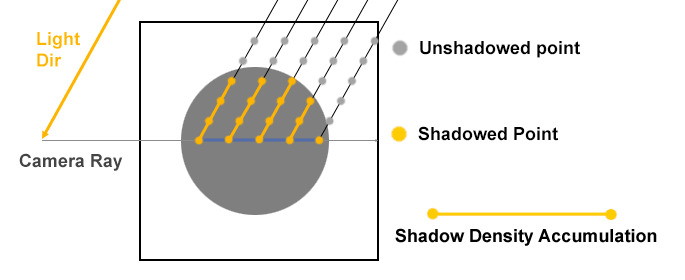}
  \caption{Illustration of secondary shadow ray marches. Source: \cite{bib:shaderBits2016}}
  \label{fig:Tracing_Shadows}
\end{figure}

To get acquainted with the Unreal Engine and to refresh some concepts of volume rendering, we followed a YouTube tutorial series \cite{bib:UE53_volume_tutorial_Ventura2023}, which is based on
the ShaderBits Blog \cite{bib:shaderBits2016}. This way, it was straight forward to get a dynamically lit, self-shadowed volume renderer up and running that is also able to receive shadows from its surrounding geometry. Additionally the volume can be "painted" at runtime, making it fully dynamic. See fig. \ref{fig:VolumeRenderingFromScratch} for an illustration.
While this approach was quickly set up, it has its limitations: On the one hand, it uses pseudo volume textures, i.e. 2D-Textures that contain tiles which represent low-resolution 2D-slices of the volume. Compared to a "true" volume texture, this restricts the effective maximum volume resolution even more.
But the most significant limitation is the lighting being recomputed for each frame, i.e. for each voxel sample taken along a ray, a secondary ray march towards the light source needs to be done, see fig. \ref{fig:Tracing_Shadows}. The performance impact is prohibitive for larger volumes. Adding a precomputed lighting cache that is only updated on voxel content or light source change would require an additional render pass and custom C++ and shader code, which is not straightforward anymore and has been done already, as we will explore in the next subsection.

\subsection{Using an existing Plugin: TBRayMarcher}
\label{subsec:tbrm}

The rendering technique of the TBRayMarcher Plugin for the Unreal Engine \cite{bib:TBRaymarcherPlugin2024} is based on an implementation by Sundén and Ropinski\cite{bib:TBRM_basePaper_Sunden}.

\begin{figure}[H]
  \centering
  \includegraphics[width=0.25\textwidth]{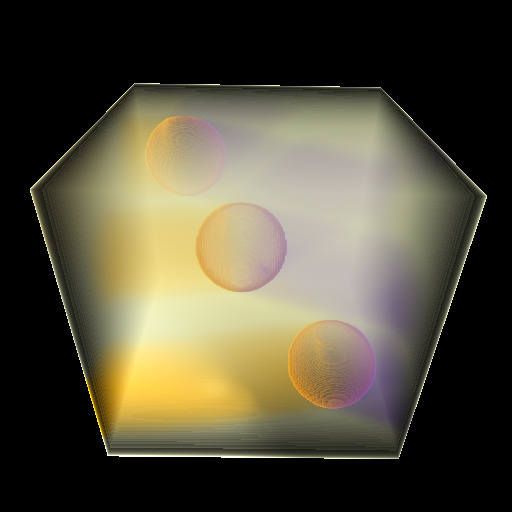}
  \caption{Illustration an illumination cache volume for one yellow and one blue point light source illuminating a group of spheres. Source: \cite{bib:TBRM_basePaper_Sunden}}
  \label{fig:Illumination_Volume}
\end{figure}

Using the simplifying assumption that the volume material scatters light isotropically (i.e., equally in all directions), the lighting distribution inside the volume can be pre-computed and
stored in an illumination cache
\footnote{which does not necessarily need do have the same extents as the volume dataset itself. This will become relevant in section \ref{subsec:svt}},
see fig. \ref{fig:Illumination_Volume}. This drastically improves rendering performance.

The plugin uses "true" volume textures (i.e. no pseudo-volume textures) and
reports to work best using DirectX 11 as a rendering backend, while the author reports issues with DirectX 12 that he has not been able to resolve yet,
and Vulkan is reported to not work at all.

We evaluated the plugin on Unreal Engine 5.3 and later migrated with its updated version to UE 5.4.

Being a plugin targeted at medical datasets, it supports importing DICOM and MHD files per drag \& drop into the engine editor. We were often unsuccessful with importing DICOM files due to the DICOM loader being outdated. We worked around that issue by using \href{https://www.slicer.org/}{Slicer}
\footnote{\url{https://www.slicer.org/}}
, a tool for various tasks around medical imaging, to load the DICOM and export an MHD file. In the meantime, the DICOM loader has been updated.

The plugin comes with a GUI allowing to select from several pre-defined transfer functions. Finding appropriate transfer functions for seismic data
is out of the scope of this work; instead we focus on overcoming certain technical limitations and explore ways to implement further tooling once we have settled on a
prospective approach.

We tried to visualize the Kolumbo 3D seismic data set (5.9 billion voxels) with this plugin.
We used Matlab with \href{https://www.mathworks.com/matlabcentral/fileexchange/53109-seislab-3-02}{Seislab 3.02}
\footnote{\url{https://www.mathworks.com/matlabcentral/fileexchange/53109-seislab-3-02}}
to load the SEG-Y file, then used Matlab's \href{https://www.mathworks.com/products/medical-imaging.html}{Medical Imaging Toolbox}
\footnote{\url{https://www.mathworks.com/products/medical-imaging.html}}
to write MHD files.
We normalized the data to fit into 8 bit precision
\footnote{The voxel data in the original Kolumbo SEG-Y file has a 32bit floating point precision and a value range of $[-34.50, +36.45]$.},
yielding a 5.5 GB MHD file.

The maximum extents along each dimension for volume textures in both DirectX 11 and DirectX 12 is 2048, so the full dataset wouldn't fit into a single texture and would require chunking into multiple volume textures.
But we even encountered crashes way below this limit of 2048 voxels per dimension. We did not investigate this any further in terms of finding the practical limit, as it may depend on the size, available free memory and and fragementation of the VRAM.
The issue can in principle resolved by using DirectX 12's reserved resources feature, but as mentioned above, this is not possible with TBRM, as it only works with DirectX 11 so far.

\begin{figure}[H]
  \centering
  \includegraphics[width=1.0\textwidth]{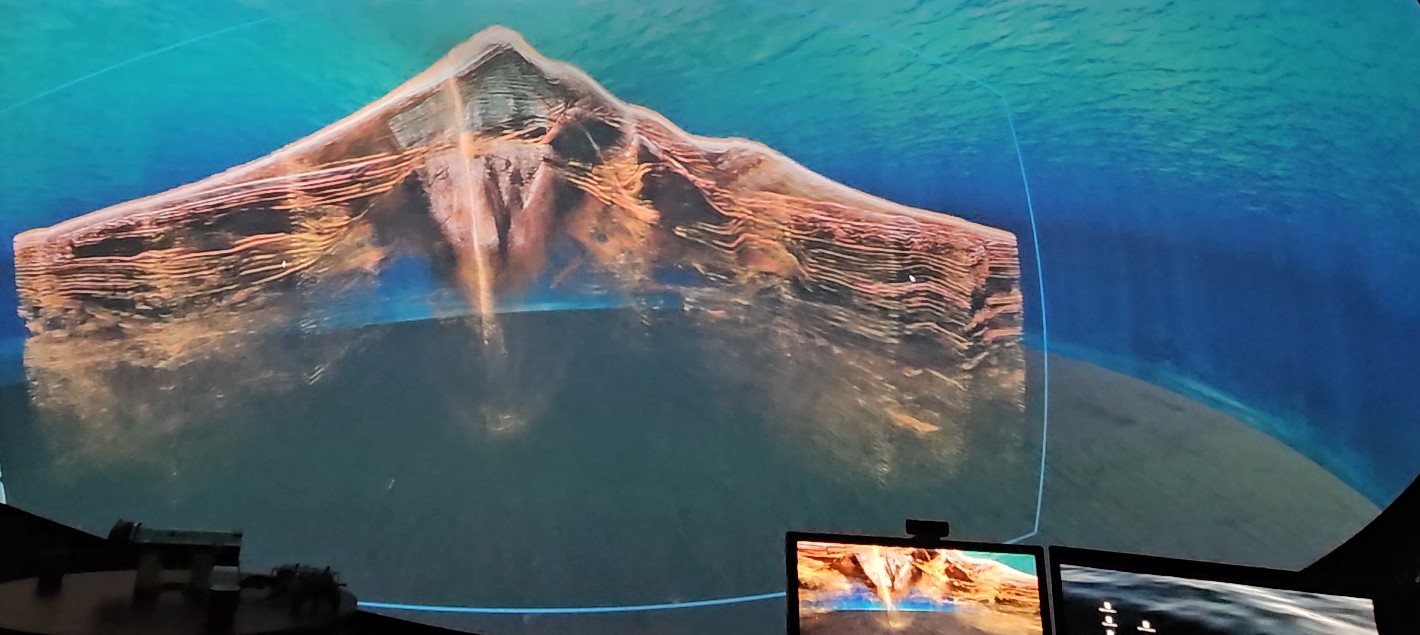}
  \caption{The Kolumbo dataset chunked into nine parts, each rendered separately using a volume texture for each chunk with the \textit{TBRayMarcher} plugin. The scene is displayed in our ARENA2 visualization dome.}
  \label{fig:Kolumbo_TBRM_Dome1}
\end{figure}

But first, we tried to get as far as possible using the TBRayMarcher:
We chunked the dataset into nine sub-volumes @ $512 \times 935 \times 1501$ voxels, loaded each separately into a \lstinline{BP_RaymarchVolume} actor and  displayed them side-by-side, see fig \ref{fig:Kolumbo_TBRM_Dome1}.

This approach allows for beautiful rendering with great performance on a single machine, at least on the RTX 3500 Ada GPU, see section \ref{sec:results}.
Here, VRAM seems to be the biggest limiting factor.

\begin{figure}[H]
  \centering
  \includegraphics[width=1.0\textwidth]{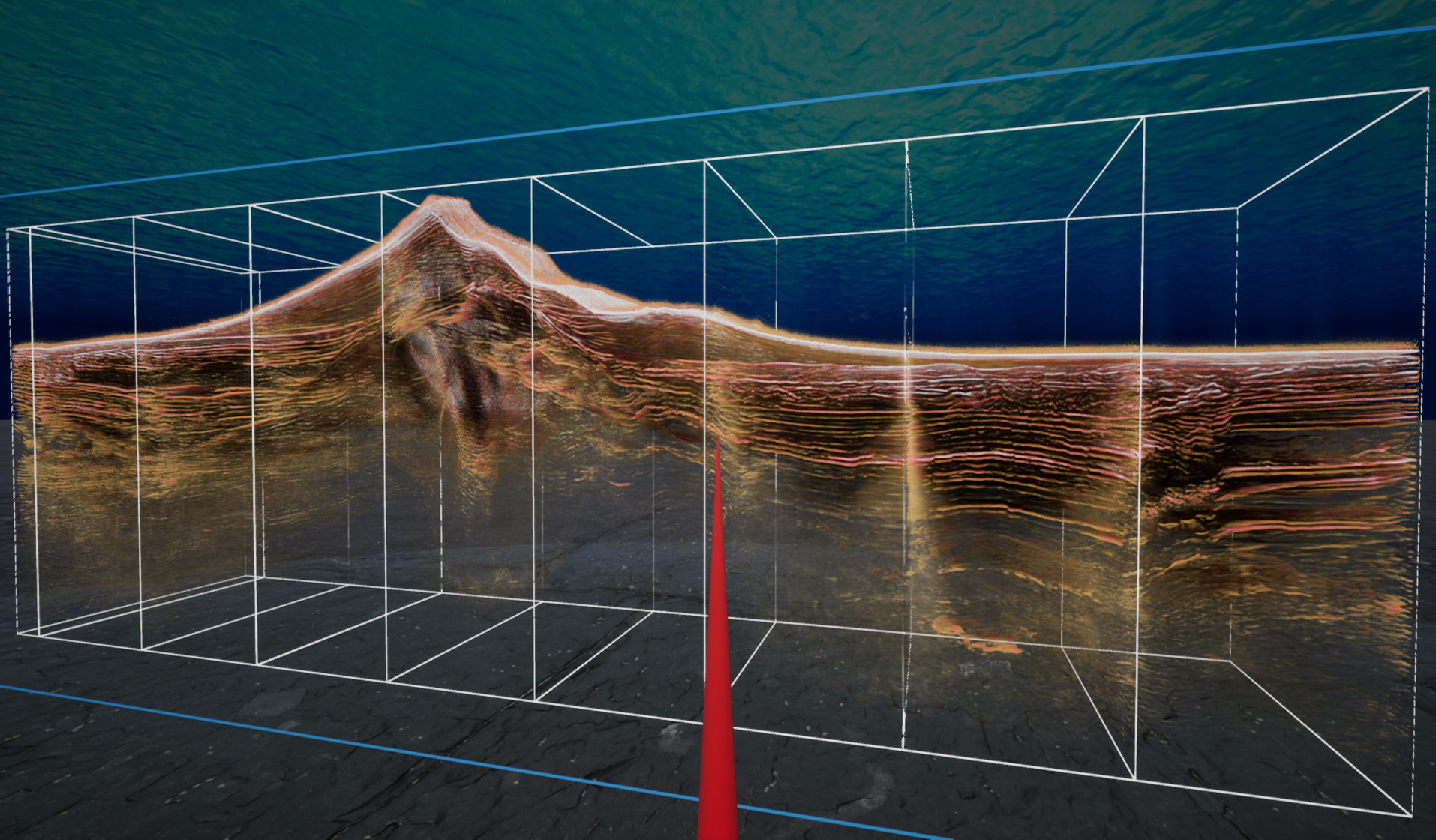}
  \caption{Illustration of the incorrect lighting at the chunk borders, empazised by showing the chunk's bounding boxes: The virtual light source is on the left, and each chunk is lit separately as if the others wouldn't exist. This makes the borders incorrectly bright.}
  \label{fig:Kolumbo_TBRM_ChunkBorders}
\end{figure}

Although this approach has some great features, it has two significant drawbacks.
On the one hand, the chunking results in incorrect lighting at the chunk borders, see fig. \ref{fig:Kolumbo_TBRM_ChunkBorders}.
On the other hand, the TBRayMarcher project is maintained basically by one person. While beging a great plugin in general, some parts are not very polished, and having to keep up with the engine updates makes the project's updates prone to regressions. This means that req. \ref{req:Sustainability} (Sustainability) is not entirely met.
This made us look for solutions that are developed and/or maintained by Epic Games,
the developers of the Unreal Engine, directly. We will investigate two approaches
in the following two subsections \ref{subsec:niagara} and \ref{subsec:svt}.

\subsection{Adapting an existing System: Niagara Fluids}
\label{subsec:niagara}

Inspired by some experiments of the Virtual Reality \& Immersive Visualization Group
at RWTH Aachen University, we adapted an Unreal-internal system that is originally intended to simulate and render both particle- and grid-based fluids: The
\href{https://dev.epicgames.com/community/learning/paths/mZ/unreal-engine-niagara-fluids}{Niagara Fluids Plugin}
\footnote{\url{https://dev.epicgames.com/community/learning/paths/mZ/unreal-engine-niagara-fluids}}
,
which at the time of writing is in beta status. It is a system with a distinct look and feel that is different from the rest of the engine, so it takes some time to get used to it, and being a beta feature, documentation was sparse at the time of writing.

The idea is simple: Grid-based simulations use - as their name suggests - regular voxel grids for both simulation and rendering. We can ignore the simulation part (a "do nothing"-simulation) and just use the rendering part.

\begin{figure}[H]
  \centering
  \includegraphics[width=1.0\textwidth]{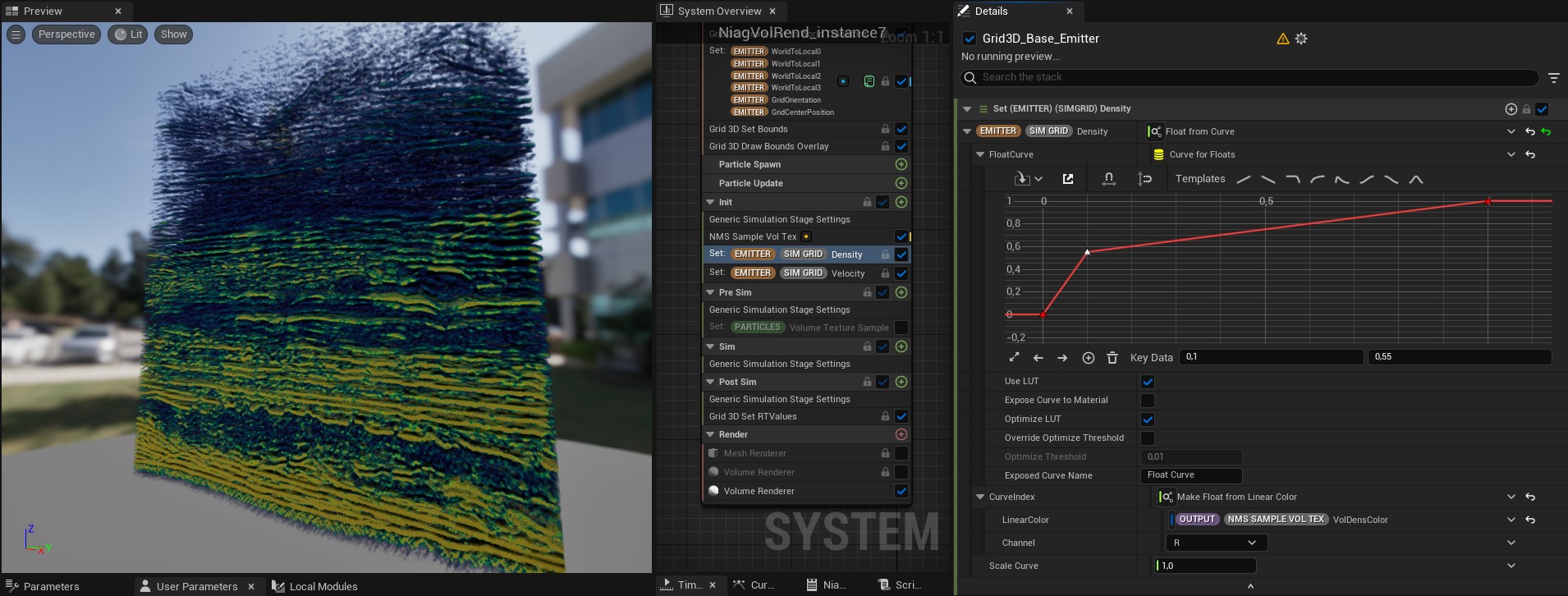}
  \includegraphics[width=1.0\textwidth]{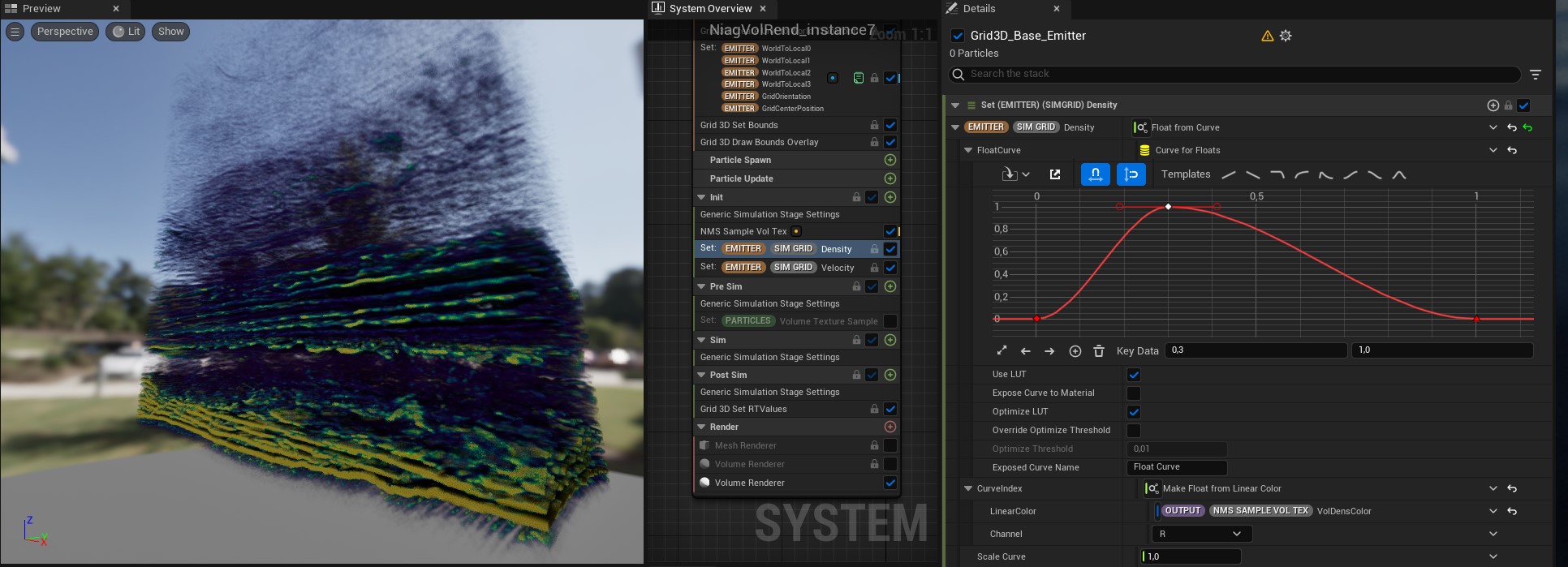}
  \caption{The Niagara system, complemented with an appropriate material, allows - in principle - to build a an authoring tool for simple transfer functions.
  }
  \label{fig:Niagara_TF}
\end{figure}

The approach sounds intriguing, because (after some setup, see below), we have an Unreal-Native approach to volume rendering (requirement \ref{req:Sustainability}, Sustainability). Additionally, we can use and modify various materials from the \textit{Volume} material domain (req. \ref{req:Flexibility}, Flexibility).
An interesting feature is that there is some built-in support for interactively authoring some kind of simple transfer function (req. \ref{req:Explorability}, Explorability), at least in the Editor (and not at runtime of the "dame"), see fig. \ref{fig:Niagara_TF}. In practice, there may be more appropriate approaches using color curves accessed in materials via texture atlas nodes.

\begin{figure}[H]
  \centering
  \includegraphics[width=1.0\textwidth]{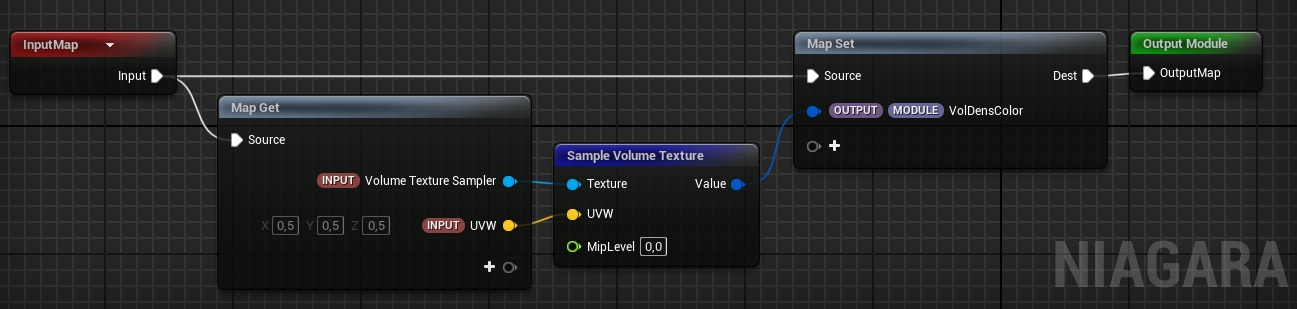}
  \caption{Niagara Module Script to sample from a volume texture.}
  \label{fig:NiagaraModuleScript_SampleVolume}
\end{figure}

First, we followed a tutorial \cite{bib:UE5_niagara_for_volume_render_tutorial_renderBucket2023}, then extended
the approach to sample volume textures (instead of 2D or pseudo volume textures).
For some reason, there was no so-called \textit{Niagara Module Script} to sample Volume textures, so we had to create it ourselves (see fig. \ref{fig:NiagaraModuleScript_SampleVolume}).

\begin{figure}[H]
  \centering
  \includegraphics[width=1.0\textwidth]{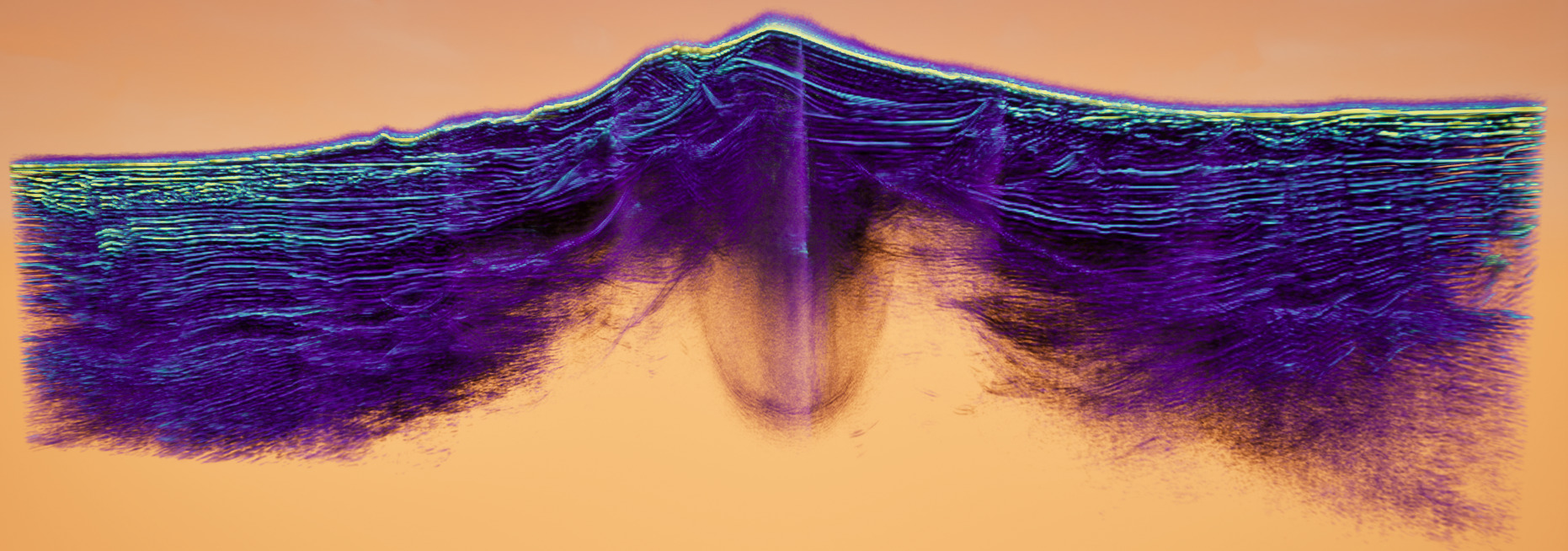}
  \caption{The Kolumbo dataset chunked into nine parts line in fig. \ref{fig:Kolumbo_TBRM_Dome1}, but each chunk is displayed as a distinct Niagara System. This results in the same lighting artifacts at the chunk's borders as with the TBRayMarcher approach. The rendering is blurry because of a low internal resolution.}
  \label{fig:Kolumbo_full_Niagara_RTX5000}
\end{figure}

We then were able to feed the Volume Textures previously imported by the MHD loader of the TBRayMarcher plugin. The results are qualitatively comparable to the TBRayMarcher-approach, see fig. \ref{fig:Kolumbo_full_Niagara_RTX5000}.

There is one severe limitation to this approach when it comes to huge datasets:
The misuse of a complex physics simulation system adds a heavy burden in terms of VRAM to the system: Even if it is not updated over time, the grid data for the simulation must be allocated on top of the texture data we want to display. To make things worse, the internal resolution of a simulation grid must be much smaller than the actual data we want to display: Even with a per-chunk resolution as small as 384 for the largest dimension
\footnote{I.e., the volume eventually displayed is about $4^3=64$ times smaller than
  the original dataset.},
we encounterd out-of-VRAM-crashes with the RTX 3500 Ada GPU (12 GB VRAM) if we try to show more than five chunks at the same time. On the RTX 5000, the full dataset fits into its 16 GB VRAM, but it is also brought to its knees when significantly increasing the internal resolution.

So, requirement \ref{req:Accuracy} (Accuracy) is not met for this approach.

There are also some limitations for the authoring of the transfer function:
\begin{enumerate}
  \item Authoring is only possible in Editor mode, not in Game mode. While there may exist ways to render Editor content to NDisplay devices (using a multi user session), it is not an official use case and may not work under all circumstances, especially when handling very large textures. \todo{ask armin for a way to test this or some documentation/tutorial.}
  \item If one wants to have more control than a density-to density mapping,
        but instead also specify color and alpha curves, the results would be written for
        each voxel, \footnote{i.e. quadrupling the memory footprint compared to a scalar density value},to be later evaluated by a ray marcher and a material. In constrast, specifying a four-channel RGBA-1D-lookup texture to a material is much more efficient.
  \item It may not be easy to create and select multiple transfer functions in this approach.
\end{enumerate}

There may or may not be other issues, like creating an appropriate inheritance hierarchy
for multiple niagara systems, so that each change is reflected for each chunk. We did not investigate this any further, as the high VRAM consumption accompanied with the low rendering resolution is prohibitive for our use case.

\subsection{Using an experimental feature: Sparse Volume Textures (SVT) / Heterogeneous Volumes}
\label{subsec:svt}

\subsubsection{Introduction to SVT}

\begin{figure}[H]
  \centering
  \includegraphics[width=1.0\textwidth]{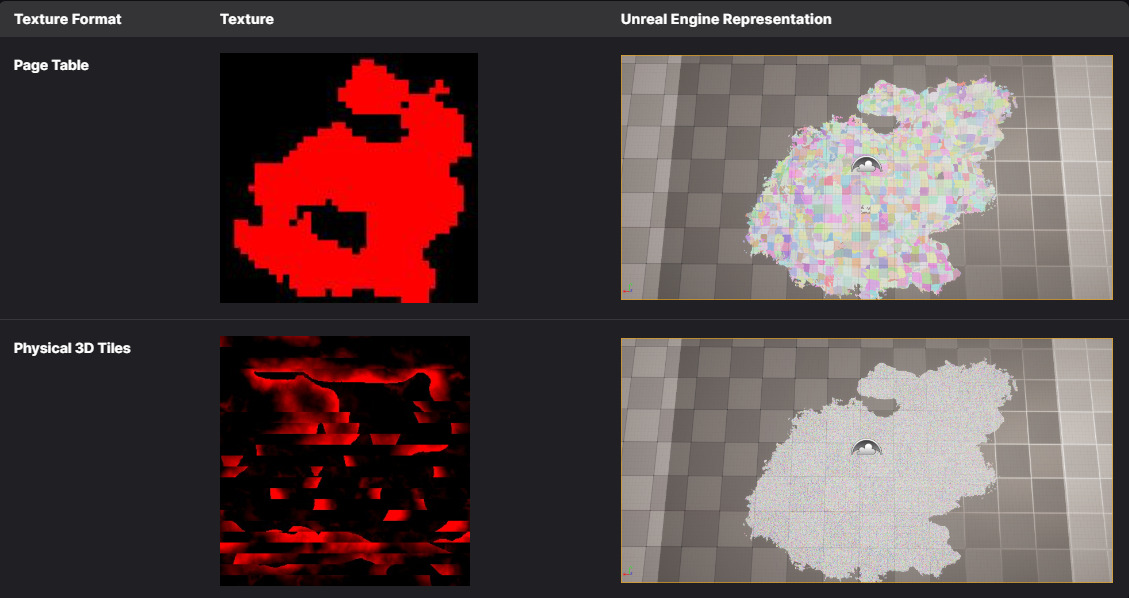}
  \caption{The page table approach of SVT: The Voxels are chunked into tiles, so that only non-empty tiles need to be stored. The lookup of the tiles happens through a volume texture that serves as a page table.
    Source: \href{https://dev.epicgames.com/documentation/en-us/unreal-engine/sparse-volume-textures-in-unreal-engine}{Epic Games}
  }
  \label{fig:SVT_Pagetable}
\end{figure}

There is another built-in (experimental) feature of Unreal Engine:
\href{https://dev.epicgames.com/documentation/en-us/unreal-engine/sparse-volume-textures-in-unreal-engine}{\textit{Sparse Volume Textures (SVT)}}
\footnote{\url{https://dev.epicgames.com/documentation/en-us/unreal-engine/sparse-volume-textures-in-unreal-engine}}
.
An
\href{https://dev.epicgames.com/community/learning/talks-and-demos/1V6r/unreal-engine-creating-visual-effects-with-niagara-fluids-sparse-volume-textures-and-heterogeneous-volumes-in-ue-unreal-fest-2023}{Introduction Video}
\footnote{\url{https://dev.epicgames.com/community/learning/talks-and-demos/1V6r/unreal-engine-creating-visual-effects-with-niagara-fluids-sparse-volume-textures-and-heterogeneous-volumes-in-ue-unreal-fest-2023}}
helped with understanding some technical details. \todo{Cite this video correctly}
The idea is relatively simple, and represents a trade-off between rendering speed and data compression using a page table approach, see fig \ref{fig:SVT_Pagetable}: The Voxel Data is chunked into 3D-Tiles and written densely into a volume texture (the "physical tile data texture"), so that only non-empty tiles need to be stored. This texture is then looked up via another volume texture that serves as a page table.

\subsubsection{Different Ways to render SVTs}

\begin{figure}
  \centering
  \begin{minipage}{0.45\textwidth}
    \subcaptionbox{SparseVolumeTextureViewer}{\includegraphics[width=\textwidth]{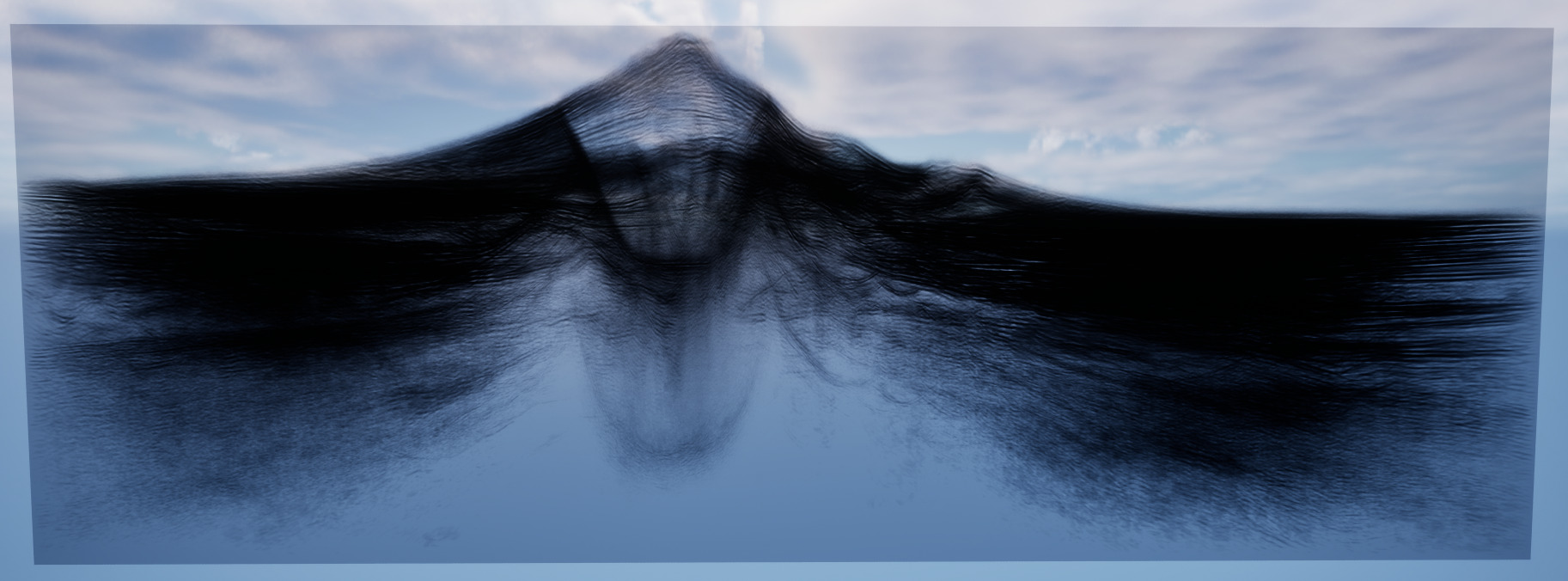}}
    \vspace{2mm} 
    \subcaptionbox{Volumetric Fog}{\includegraphics[width=\textwidth]{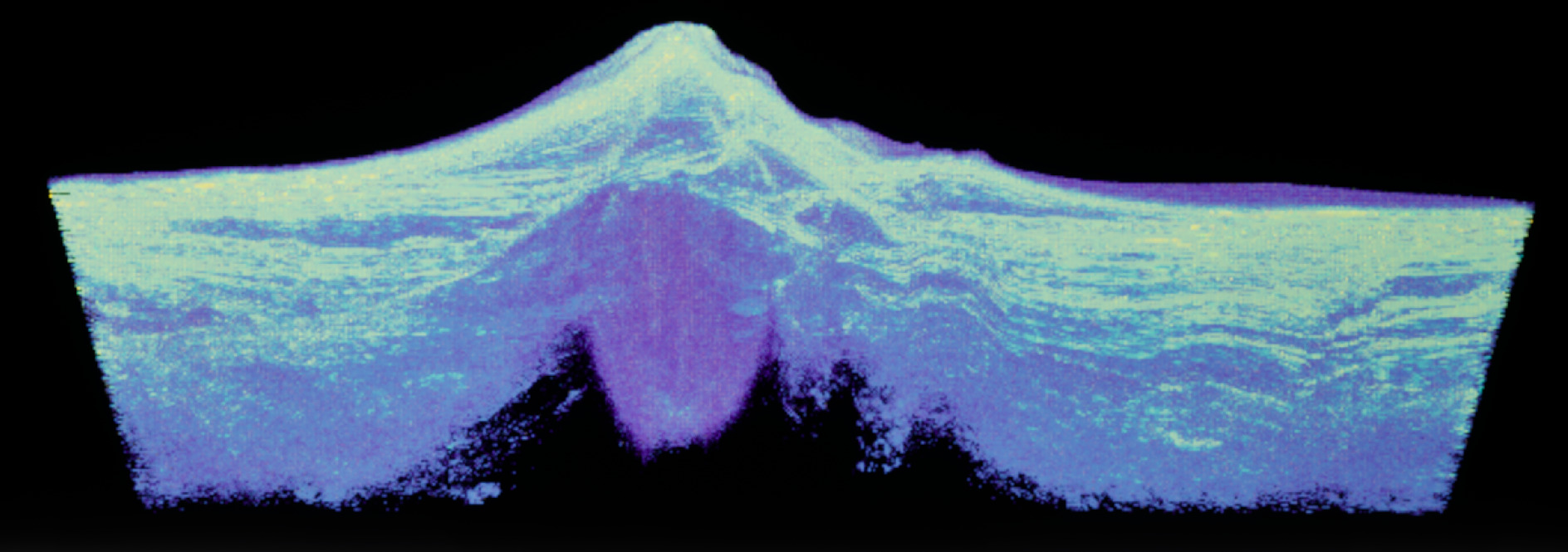}}
    \vspace{2mm}
    \subcaptionbox{\textbf{Heterogeneous Volume}}{\includegraphics[width=\textwidth]{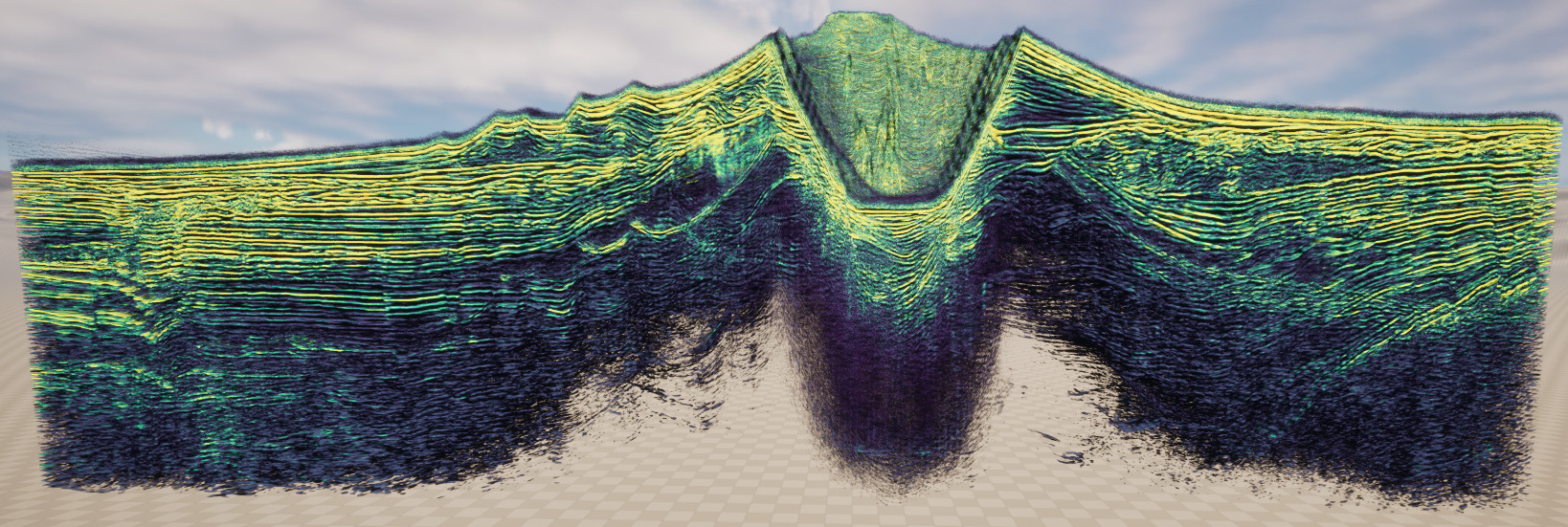}}
  \end{minipage}
  \hfill
  \begin{minipage}{0.45\textwidth}
    \subcaptionbox{Volumetric Cloud}{\includegraphics[width=\textwidth]{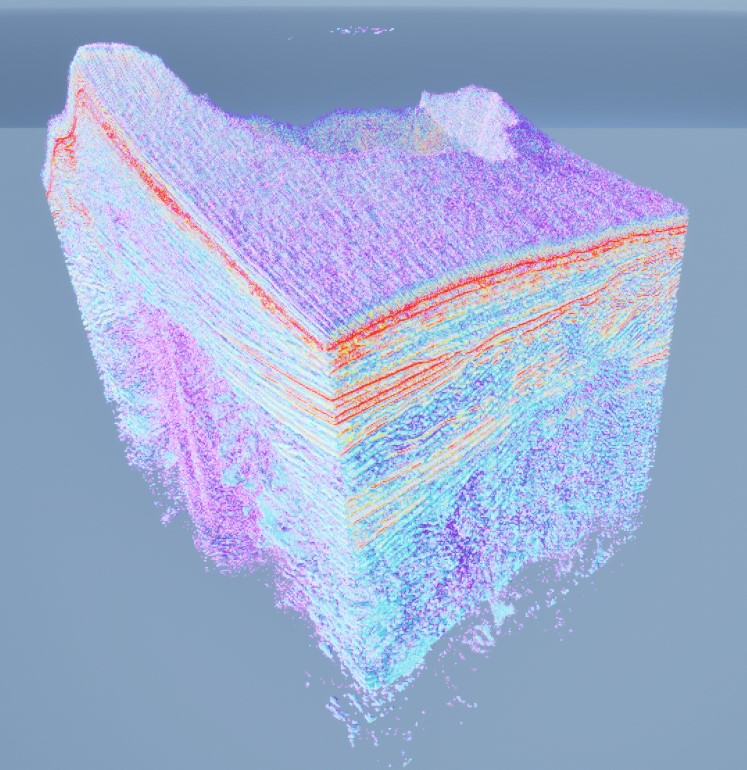}}
  \end{minipage}
  \caption{Overview of the different rendering techniques for SVT's, illustrated by the Kolumbo dataset. Only Heterogeneous Volumes (c)
    fit our needs. the SVT Viewer (a) is a debug renderer. The Volumetric Fog (b) and Cloud (d) required some tweaks to increase internal resolution and reduce noise, some taking a heavy performance hit, yet still look blurry or noisy. This is appropriate for fog and clouds, but not for visualizing scientific data.}
  \label{fig:FourWaysForRenderingSVTs}
\end{figure}

There are four distinct ways an SVT can be rendered in Unreal Engine, see fig. \ref{fig:FourWaysForRenderingSVTs}. Three of them sound inappropriate and were confirmed as misfits for our purposes, but for the sake of completeness and for the reader's amusement, we list a short description, and some example renderings are shown in fig. \ref{fig:FourWaysForRenderingSVTs}.

\begin{itemize}
  \item SparseVolumeTextureViewer: It is mostly used for debug purposes, as it can only render black extinction without any lighting, emission or color.
  \item Volumetric Fog: A Fog Material can be applied to a dummy Mesh, e.g. a cube,
        which samples the SVT. The results are blurry and ghostly, as one would expect from something that is supposed to look like fog.
  \item Volumetric Cloud: The Volumetric cloud system can sample SVTs.
  \item \textbf{Heterogeneous Volume}: This is the most appropriate way to render large general purpose data, and we settled for this technique.
\end{itemize}

\subsubsection{Importing SVT via OpenVDB}

Unreal Engine's interface to SVTs is an importer of \href{https://www.openvdb.org/}{OpenVDB}
\footnote{\url{https://www.openvdb.org/}}
files. OpenVDB stores voxel data in a sparse manner using hierarchical data structures.

We performed the conversion from the SEG-Y format to VDB via a Jupyter Notebook using \href{https://pypi.org/project/segyio/}{segyio}
\footnote{\url{https://pypi.org/project/segyio/}}
and
\href{https://www.openvdb.org/documentation/doxygen/python.html}{OpenVDB's Python bindings}
\footnote{\url{https://www.openvdb.org/documentation/doxygen/python.html}}

Upon loading the 4.4GB VDB file of the full Kolumbo dataset using 8 bit precision per voxel value, we encountered a crash in the Unreal Engine, see fig. \ref{fig:UE_Crash_VDB_import}.
We found that the import went smoothly only when the data was yet again chunked, and this time, at least four chunks were required. \newline
This is an improvement to requiring nine chunks with the previous approaches, but still far behind our expectations: No official limits were violated, and there were no principal technical limitations that would forbid loading and rendering the entire dataset.
In the end, the crash turned out to be caused by at least two (signed) \lstinline{int32} or (unsigned) \lstinline{uint32} overflows that we were able to fix; see Section \ref{subsubsec:svt:fixing}. But before diving deep into the Engine's source code, we tried a workaround, described in the next section.

\subsubsection{Workaround: Create a 4-in-1 Material}
\label{subsec:svt:workaround}

\begin{figure}[H]
  \centering
  \includegraphics[width=1.0\textwidth]{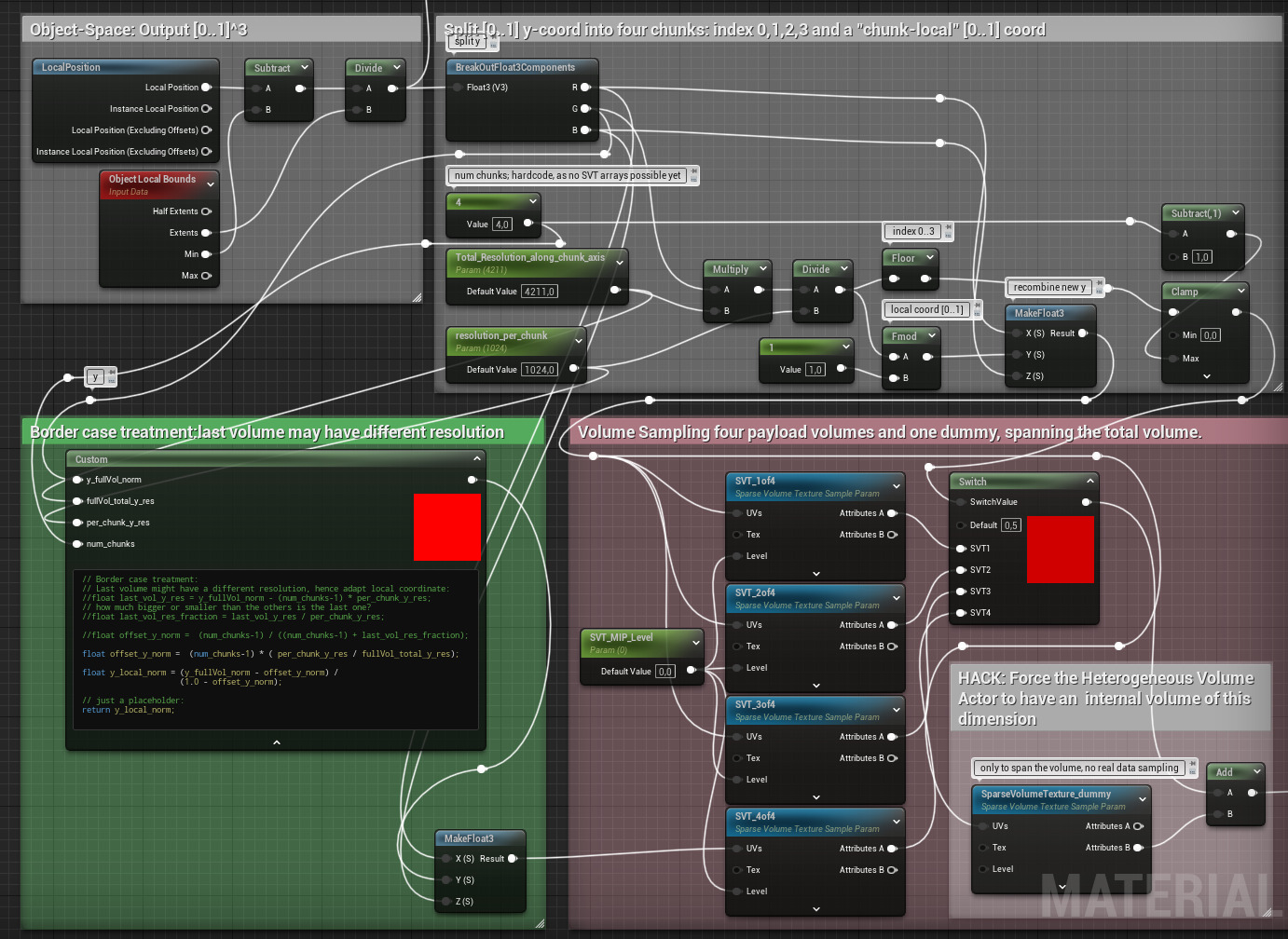}
  \caption{Workaround for lighting artifacts due to chunking: Part of a modified Material of a Heterogeneous Volume Actor (HVA): The HVA beliefs having one volume of a certain resolution. For this, we use an SVT of full resolution which spans the volume by having values only at the min. and max. corners. Then, the four payload chunks are sampled as if they were one texture. }
  \label{fig:SVT_4Chunks1Material_Hack_Material}
\end{figure}

So far, we had been unable to get rid of lighting artifacts due to chunking.
So, first, we tried a trick: Even if we are unable to \textit{import} the dataset as a whole,
we can still try \textit{rendering} it as a whole by hacking the part of the Material where the (single) SVT is evaluated, and extend it to select and sample from the four different chunks,
see fig \ref{fig:SVT_4Chunks1Material_Hack_Material}.

\begin{figure}
  \centering
  \includegraphics[width=0.75\textwidth]{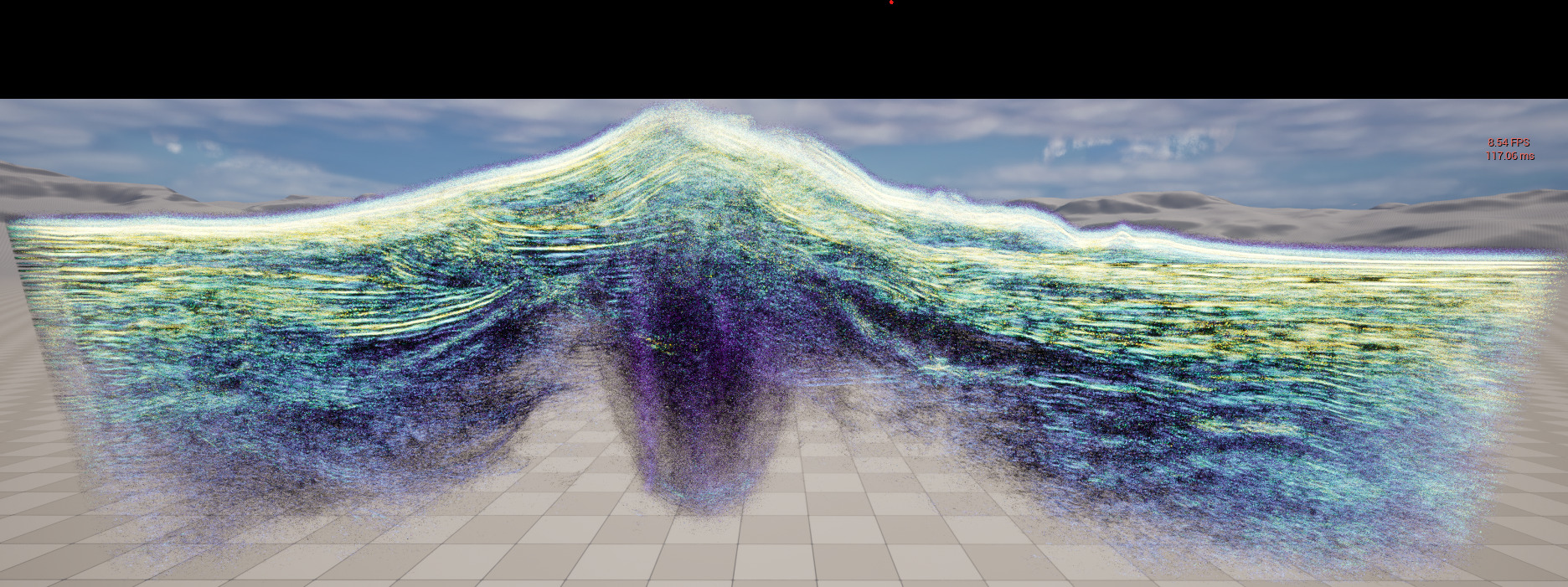}
  \includegraphics[width=0.75\textwidth]{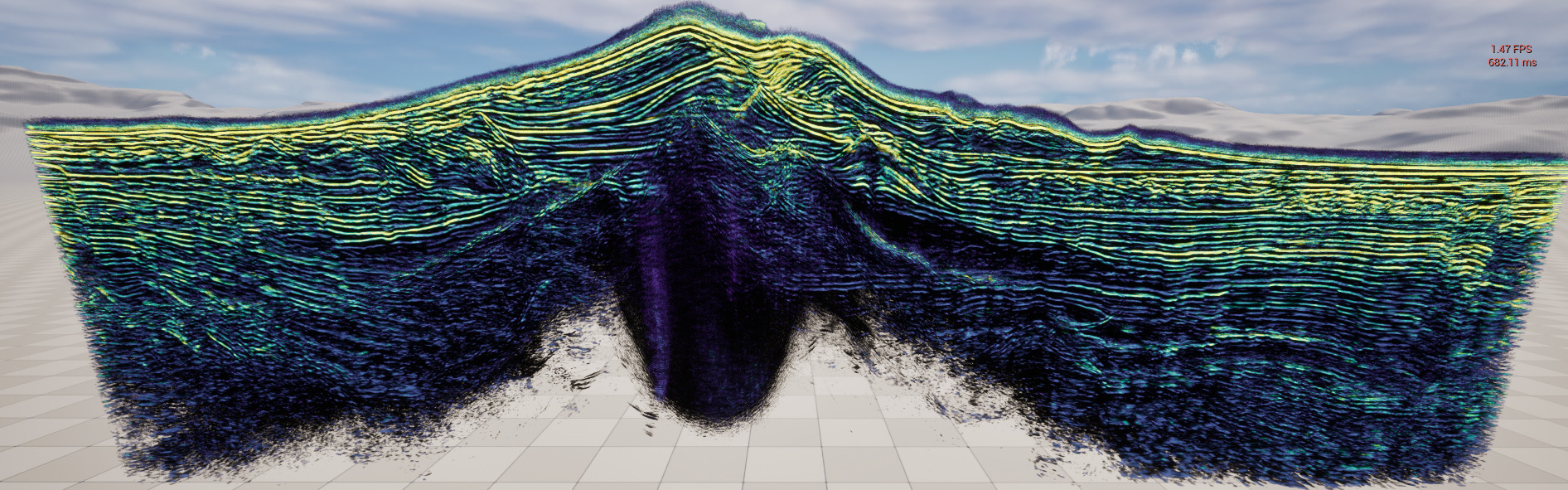}
  \caption{The workaround of creating a 4-chunks-in-1-actor material enables seamless lighting throughput the volume, at the price of a heavy performance hit. We compensate by allowing the user to navigate at a lower sampling rate: In a blueprint, we trigger the console command \lstinline{r.HeterogeneousVolumes.MaxStepCount=32} for interactivity (top) and reset is back to 512 for higher quality (bottom).}
  \label{fig:SVT_4Chunks1Material_Hack_Rendering}
\end{figure}

This way, the internal lighting cache of the Heterogeneous Volume Actor (HVA) is fed with values of all four chunks, creating a seamless rendering, see fig. \ref{fig:SVT_4Chunks1Material_Hack_Rendering}. But this trick comes at a heavy performance impact, dropping framerates to $\sim $ 1fps and forced us to the next workaround: A preview mode similar to interactivley navigating path traced scenes: We allow the user to navigate interactively using fewer samples per ray march (e.g. 32), and render with more samples (e.g. 512) when an interesting view is found.

So, we eventually managed to display the whole Kolumbo dataset interactively and without artifacts. But the cost was a complicated chunking and importing, the setup of a custom material and a huge performance hit. We did not want to settle for that, and looked for a way to fix that import crash, which we will treat in the next section.

\subsubsection{Increasing the Engine's limits}
\label{subsubsec:svt:fixing}

\paragraph{Fixing the VDB import}

\begin{figure}
    \centering
    \includegraphics[width=1.0\textwidth]{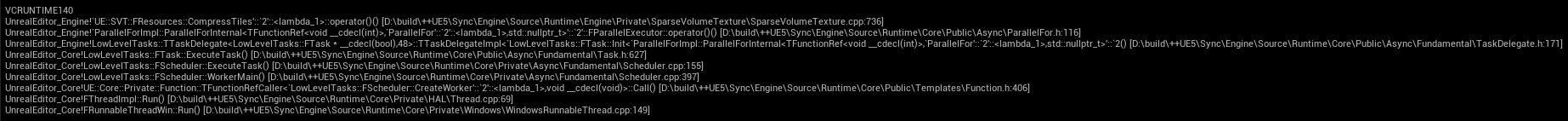}
    \caption{The crash report of Unreal Engine when loading a 4.4GB VDB file of the full Kolumbo dataset using 8 bit precision.}
    \label{fig:UE_Crash_VDB_import}
  \end{figure}

When we first tried to import the Kolumbo dataset as a 4.4 GB file VBD file with 8 bit precision per voxel, we encountered a crash of the Unreal Engine,  both in version 5.3 and 5.4, see fig. \ref{fig:UE_Crash_VDB_import}.
\footnote{Update: We tested the import later in UE 5.5.4, but the issue persists.}.
There were no official limits we were exceeding, so we investigated.

Although we found a workaround for the unsuccessful import, it was complicated, hard-coded to the number of chunks and resulted in very poor performance. The crash reported an access violation in the \lstinline{SVT::FResources::CompressTiles(...)} function. Given that the VDB file is larger than $2^{32}=4GB$ and knowing that GPU related code tends to omit 64 bit types for performance reasons
  \footnote{GPUs have significantly fewer computation units for 64 bit integer and floating point arithmetic, severely reducing the parallelism, which is the main reason to use a GPU in the first place.},
  we suspected a 32-Bit integer overflow and had a look at the source code:

  We noticed that the CPU-side representation of the dataset (both for storage and in RAM) uses an internal compression scheme using occupancy bits. This allows to only store non-empty voxels, of which there are about 2.6 billion in the dataset. This is within the range of unsigned 32bit integer (\lstinline{uint32}, $2^{32}=4G$), but outside the range of \textit{signed} integer (\lstinline{int32}), which has only 31 bits to represent a positive number ($2^{31}=2G$). Indeed, upon changing some index types from \lstinline{int32} to \lstinline{uint32}, the dataset was completely imported.

  \paragraph{Fixing the GPU-Upload}

\begin{figure}
  \centering
  \includegraphics[width=1.0\textwidth]{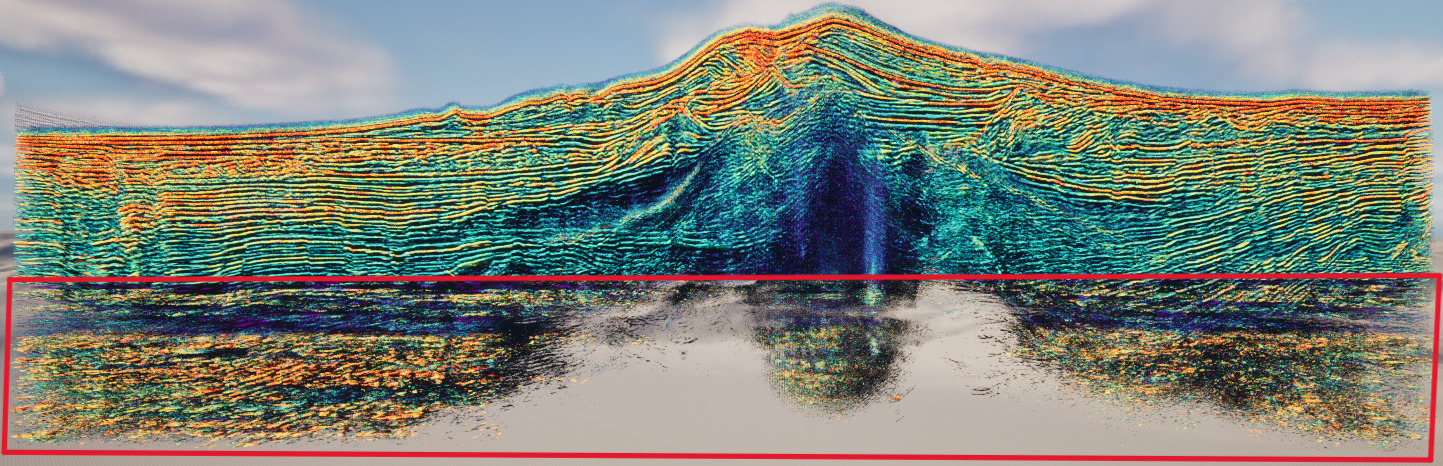}
  \includegraphics[width=1.0\textwidth]{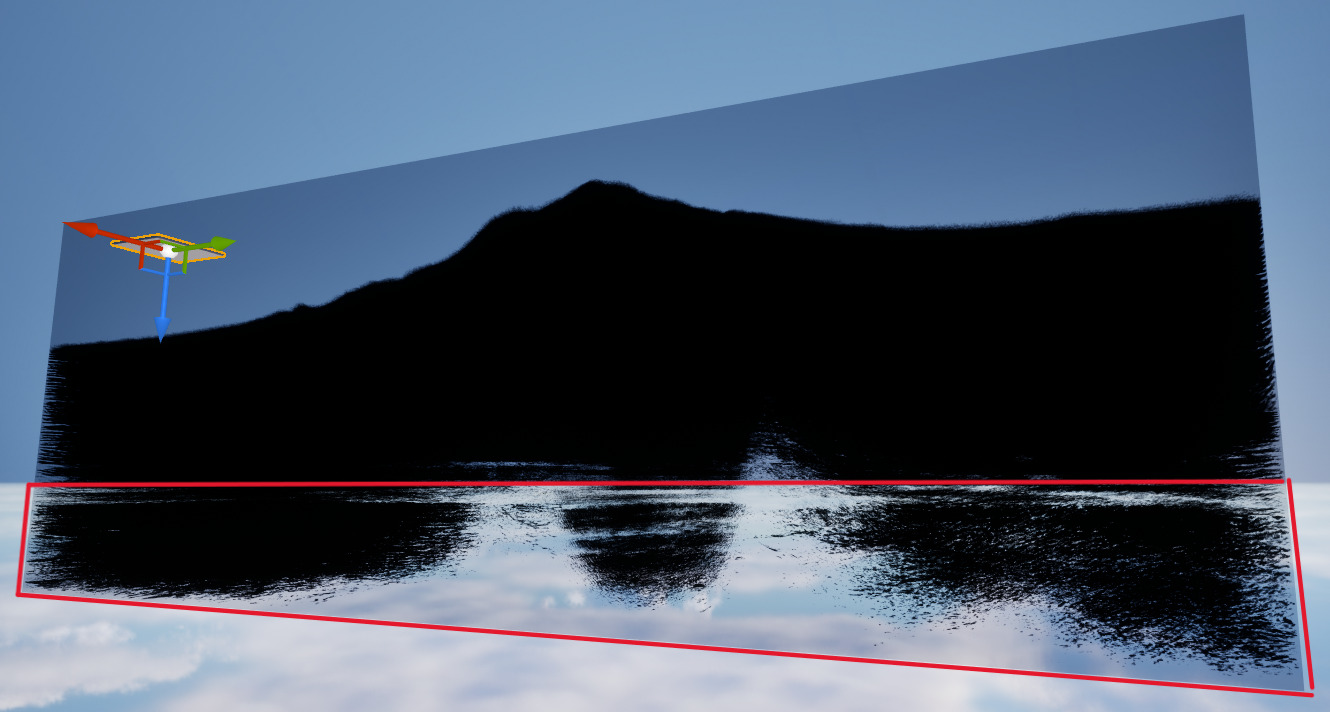}
  \caption{
    The rendering of the lower parts of the Kolumbo dataset was incorrect initially, both with the \textit{Heterogeneous Volume Actor}  (top) and the
    minimalist debug renderer (\textit{Sparse Volume Texture Viewer}). Note that the tiling mechanism makes the incorrect values appear roughly at plausible geometric tile locations (the tile index and occupancy buffers do not overflow), but the values within the lower tiles were still incorrect.
  }
  \label{fig:SVT_issue}
\end{figure}

  While the dataset was now loaded without a crash and we were able to render it with decent performance (about 10 fps), the lower parts were displayed incorrectly, see fig. \ref{fig:SVT_issue}. This looked like another \lstinline{int32}  overflow issue: Of 2.6 billion voxels, $2^{31}=2.1$ billion may have been uploaded correctly, leaving about one fifth of them incorrectly indexed. The suspicion hardened when even the simple debug draw utility, the \textit{SparseVolumeTextureViewer} had the same issues (see fig. \ref{fig:SVT_issue}, bottom): An inspection of its rendering shader (\lstinline|VisualizeSparseVolumeTexture.usf|) showed no signs of an issue on the display side.

The compute shader to upload and decompress the data into a Volume Texture on the GPU (\lstinline{UpdateSparseVolumeTexture.usf}) uses only \lstinline{uint32} integer types, suggesting to support 4GB upload buffers on the GPU side
\footnote{
  Our dataset's upload buffer comes close to $2^{32}=4$GB, but does not exceed it: while the compressed upload buffer has $2.6 \times 10^9$ Bytes, its $16^3$ tiles are padded
  to $18^3$ voxels, in order to enable correct trilinear interpolation at the tile's borders. On average, this increases their size of the upload buffer by a factor of $(18/16)^3$ = 1.424.
  On top, the upload buffer also contains all 3D Mip levels, enlarging it by another factor of $\sum_{n=0}^{\infty } (1/2^n)^3 \to 1.143$.
  The total upload buffer size is about $ 2.6 \times 10^9 \times 1.424 \times 1.143 = 4.2318432 \times 10^9$ Bytes, which is still slightly less than $2^{32}=4.294 \times 10^9$ Bytes.
}
, so we expected the overflow on the CPU side.

Eventually, further debugging enabled us to identify some integer types that caused some overflows: Not only the upload-buffer calculations were affected,
but also one signature of Unreal's Render Hardware Interface (RHI) had its return type to be modified: \lstinline{virtual uint64 FDynamicRHI::RHIComputeMemorySize(FRHITexture* TextureRHI)}
\footnote{It is not clear to us if this function's return value is for statistical purposes only, or if it was another source of rendering issues.}
: The volume texture to hold all the tile data comes close to its limits ($2048^3$), where the reserved size is reported to be $6.7 \times 10^9$ bytes.

In the end, we were able to render the Kolumbo dataset without artifacts or incorrect values at about 20 fps, cmp. fig. \ref{fig:Kolumbo_HetVol_full_fixed}.

\subsection{Alternatives to Unreal Engine: ParaView and the NVIDIA IndeX Plugin}
\label{subsec:paraview_index}

\begin{figure}[H]
  \centering
  \includegraphics[width=0.6\textwidth]{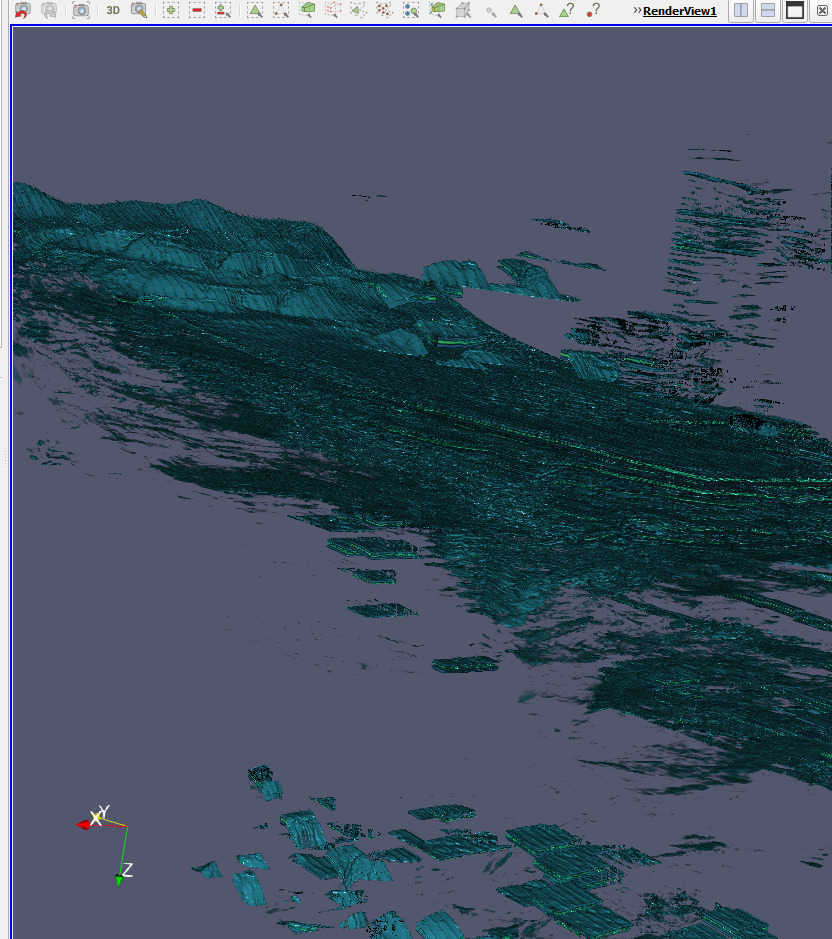}
  \caption{Popping/Jumping artifacts we encountered with Paraview's IndeX plugin if the data size approaches the GPU's VRAM. They look like an improperly sampled volume texture that serves as a cache for streamed-in tiles.}
  \label{fig:IndeX_artifacts}
\end{figure}

As mentioned in Section \ref{sec:relatedwork}, the combination of ParaView with NVIDIA IndeX sounds like a promising approach for bringing huge volumetric datasets into the dome.
We experimented with a single-node approach, i.e. where no multi-GPU clusters are leveraged for distributed rendering of one single frame. The reason is that we need our rendering cluster to render the respective perspective for its dome projector, so there are no capabilities left to concentrate the cluster on rendering portions of just one frame. We chunked the data with its original \lstinline|float32| precision into several.vti files
\footnote{As the IndeX plugin expects point data (and not cell data), the chunk data has to overlap on the border slices, otherwise, no interpolation would be possible.}
and combined them into a.pvti file ("p" is for "parallel", in the sense that the chunks can be handled by distributing them among the render nodes).
Unfortunately, we encountered multiple issues with this single node approach as soon as the total size of the data set approached the VRAM size of 12 GB:
\begin{enumerate}
  \item With increasing data set size, we encountered more and more popping/jumping rendering artifacts that look like an improperly used caching mechanism for out-of-core rendering, see fig. \ref{fig:IndeX_artifacts}.
  \item Increasing the data set size even further, Paraview hangs up upon selecting the "NVIDIA Index" representation.
  \item Still increasing the data set size, ParaView already creashes on attempting to lad the dataset (i.e. this seems unrealted to the IndeX Plugin.)
\end{enumerate}

We are aware that we tried to misuse the software here, which clearly has a different use case. In the long run, we still consider this approach worth pursuing, but then we have to embrace the complexity and additional cost of cloud-based remote rendering. It is not entirely clear to us if the two use cases of ParaView's and IndeX's cluster capabilities are possible to combine: We have not yet explored if ParaView allows at the same time for both clustered multichannel rendering (enabling dome rendering) and clustered rendering of the whole volume data set for each dome-channel node. This would imply a "cluster of clusters" setup, introducing a whole new set of technical challenges.
These are very interesting technical questions, and we would like to explore these possibilities in the future, but for at the time of writing we concentrated on solutions using only hardware available on the premises.

\subsection{Not Using Unreal Engine: scenery/sciview}
\label{subsec:sciview}

We encountered scenery/sciview \cite{bib:scenery/sciview} only later in our evaluation, when we already had a strong focus on the Unreal Engine. The prospect of CAVE-support
\footnote{
  CAVE's usually do not need warping and blending functionality, which is mandatory in domes. Nevertheless, our calibration provider
  \href{https://vioso.com/}{VIOSO}
  (\url{https://vioso.com/})
  offers a feature called \href{https://helpdesk.vioso.com/documentation/vioso6-using-calibrations/desktop-hooking-anyblend-technology/}{Desktop Hooking}, allowing to apply the warping and blending directly to a Windows desktop and hence all applications running on it. This way, at least for monoscopic scenarios without head-tracked update of the virtual camera's pose, only an appropriate view-frustum and its pose must be provided to an application, while it can stay ignorant of the warping and blending.
},
out-of-core volume rendering and a tool set specifically tailored for scientific visualization, however, made us experiment with the sciview Fiji plugin. We wanted to get an quick idea of the out-of-core rendering performance without engaging with the project on a source code level prematurely (some relevant documentation is quite sparse or non-existing).

However, we encountered some issues. This may have to do with some particularities
involving incompatible Java versions or some restrictions of the plugin, which may not reflect all capabilities of the actual rendering backend (called \textit{scenery}).

In summary, scenery/sciview does not currently meet requirement \ref{req:Compatibility}: compatibility and synergy with existing projects of ARENA2, and requirement \ref{req:Accessibility} (accessibility) is t.b.d.: Getting acquainted with Kotlin and the Java-based ecosystem, which so far had not been strongly associated with advanced real time computer graphics, adds to the learning curve. Nevertheless, we plan to evaluate the project more thoroughly in the near future.


\section{Results and Discussion}
\label{sec:results}

We summarize our results in table \ref{table:results_comparison} and give an estimate on how much the different approaches meet the requirements specified in section \ref{sec:requirements}.
Concerning the most requirements, the different approaches do not diverge that much from one another, as it is expected for plugins to a software that is growing more and more into a general purpose solution for various use cases. The most prominent differences are supported data sizes, display accuracy and rendering performance. We plan to contribute our patch for the SVT importing to the Unreal Engine; until then, requirements like wide adoption, sustainability, an accessibility are not met for the modified Engine.

\todo{check on master the per-chunk-performance of modded SVT w/  lighting downsample 4 and 1024 max samples}

\begin{table}[h!]
  \begin{threeparttable}

    \begin{tabularx}{\textwidth}{X|XXXX}
      \cline{2-5}
                                                         & \multicolumn{1}{X|}{TBRM}                      & \multicolumn{1}{l|}{Niagara}               & \multicolumn{1}{l|}{Het. Vol. (vanilla)}
                                                         & \multicolumn{1}{l|}{Het. Vol. (modified)}                                                                                                                            \\ \hline
      \multicolumn{1}{|l|}{Display size limits
      (GV/chunk)}                                        & $\approx 1$                                    & $\approx 0.45$ \tnote{b}                   & $\approx 1.5$ \tnote{e}                  & \textbf{$\geq 6$} \tnote{e} \\ \cline{1-1}
      \multicolumn{1}{|l|}{Display size limits
      (GV total)}                                        & $> 6$                                          & $\approx 0.28$ \tnote{c}                   & $\geq 6$  \tnote{e}                      & $\geq 6$ \tnote{e}          \\ \cline{1-1}
      \multicolumn{1}{|l|}{Seamless full Kolumbo render} & \xmark                                         & \xmark                                     & \omark \tnote{f}                         & \cmark                      \\ \cline{1-1}
      \multicolumn{1}{|l|}{Display Accuracy \tnote{1}}   & \cmark                                         & \omark \tnote{d}                           & \cmark                                   & \cmark                      \\ \cline{1-1}
      \multicolumn{1}{|l|}{Performance (per chunk)\tnote{2}}
                                                         & \cmark (>60 fps)                               & \omark (20 fps) \tnote{d}                  & \cmark (23 fps)                          & \omark (10 fps)  \tnote{h}  \\ \cline{1-1}
      \multicolumn{1}{|l|}{Performance (total)}          & \cmark (>20 fps)\tnote{a}                      & \omark (20 fps) \tnote{d}                  & \xmark (1 fps) \tnote{g}                 & \omark (10 fps)  \tnote{h}  \\ \cline{1-1}
      \multicolumn{1}{|l|}{Wide Adoption}                & \cmark                                         & \cmark                                     & \cmark                                   & \xmark \tnote{i}            \\ \cline{1-1}
      \multicolumn{1}{|l|}{Large Feature Set}            & \omark                                         & \omark                                     & \omark                                   & \omark                      \\ \cline{1-1}
      \multicolumn{1}{|l|}{Accessibility}                & \omark                                         & \omark                                     & \cmark                                   & \omark \tnote{i}            \\ \cline{1-1}
      \multicolumn{1}{|l|}{Extensibility/Adaptability}   & \omark                                         & \omark                                     & \omark                                   & \omark                      \\ \cline{1-1}
      \multicolumn{1}{|l|}{Flexibility}                  & \omark                                         & \cmark                                     & \omark                                   & \omark                      \\ \cline{1-1}
      \multicolumn{1}{|l|}{Compatibility}                & \omark                                         & \cmark                                     & \cmark                                   & \omark \tnote{i}            \\ \cline{1-1}
      \multicolumn{1}{|l|}{Explorability}                & \cmark \tnote{k}                               & \cmark \tnote{l}                           & \cmark \tnote{l}                         & \cmark \tnote{l}            \\ \cline{1-1}
      \multicolumn{1}{|l|}{Sustainability}               & \omark                                         & \cmark                                     & \cmark                                   & \omark \tnote{i}            \\ \cline{1-1}
      \multicolumn{1}{|l|}{Use case}                     & Medical datasets < 1 GV                        & Small datasets, input to fluid simulations
                                                         & general purpose datasets $\leq 1.5$ GV
                                                         & general purpose datasets $\leq 6$ GV \tnote{e}                                                                                                                       \\ \cline{1-1}
    \end{tabularx}

    \begin{tablenotes}
      \item[1] Not accounting for chunk border lighting artifacts.
      \item[2] We take the biggest possible chunk that renders without crash at interactive framerates.
      \item[a] @ WQXGA on RTX 5000; Performance is higher on RTX 3500 Ada Laptop GPU ($\approx$ 35 fps).
      \item[b] $\approx 0.768^3$; Performance is poor, and tested only for a single chunk, i.e. no side-by-side display.
      \item[c] $\approx 0.384^3$ * 5 chunks; For more chunks, UE crashes with 12GB VRAM.
      \item[d] The approach suffers less from low fps but from low data display resolution and high VRAM requirements due to severe simulation overhead.
      \item[e] Note that for sparse volume textures, the actual voxel count fitting into VRAM memory varies depending on tile size and  the number and distribution of non-empty voxel values.
      \item[f] Enabled by 4-in-1 material workaround, cmp. section \ref{subsec:svt:workaround}.
      \item[g] Huge performance hit due to workaround.
      \item[h] On RTX 5000; On RTX3500, it performs at $\approx$ 20 fps. Note that there is no significant performance drop w.r.t. dataset size, i.e. between a single 1.5 GV chunk and the full 6gV dataset.
      \item[i] Adoption, accessibility and sustainability depends on the contribution of the patch being accepted by Epic Games.
      \item[k] A cut plane mechanism and several pre-authored transfer functions for medical visualization are provided, that can be changed and whose ranges can be fine tuned at runtime.
      \item[l] Unreal Engine's material system allows for runtime manipulation of material parameters using Dynamic Material Instances (DMI) and Material Parameter Collections (MPC). This enables effects like scaling density, emission and extinction values, only showing a certain density range, mapping a transfer function to different density ranges, and exchanging the transfer function at runtime by changing a "Curve Atlas Row Parameter" node in a DMI. Unfortunately, editing the color curves themselves visually is only possible in-editor, not at runtime.
    \end{tablenotes}

  \end{threeparttable}

  \caption{Comparison of the evaluated approaches to volume rendering in UE5, using the Kolumbo dataset as an example: data size limits in gigavoxels (GV) 8bit/voxel, performance, accuracy, relevant use cases and
    an assessment of the extent the requirements of section \ref{sec:requirements} are met. \newline
    Legend: GV: gigavoxel; \cmark: Fully met; \omark: Partially met / met under conditions ; \xmark: Not met.
  }

  \label{table:results_comparison}
\end{table}

While each of the approaches has its valid use cases, we decided that the \lstinline|Heterogeneous Volume| approach with the modified engine currently meets our requirements best:
The engine modification enables us to render a sparse dataset of 5.9 gigavoxels with 8 bit precision per voxel as a whole at about 10 fps on an RTX 5000 and about 20 fps on an RTX 3500 Ada Generation Laptop GPU.
Lighting is received from an arbitrary number and type of lights in the scene. The lighting is precomputed in the lighting cache, and hence type and number of lights have no impact on rendering performance.
The volume casts shadows on itself, and both casts to
\footnote{
  We noticed some flickering artifacts when the "cast shadow" property was enabled (in project settings and the Het. Vol. Actor), which can be reduced, but not eliminated by increasing the \lstinline|Shadow Bias Factor|, so we disabled this feature.
}
and receives shadowing from its environment.

We found that for this $4211 \times 935 \times 1501$ volume,
a downsample factor of 4 and
a maximum step count of 1024 per ray
gives a good trade-off between rendering performance and quality
\footnote{Type
  \lstinline|r.HeterogeneousVolumes.LightingCache.DownsampleFactor = 4| and \newline
  \lstinline|r.HeterogeneousVolumes.MaxStepCount = 1024|
  in the console or in the \lstinline|DefaultEngine.ini| file.
}.

\paragraph{Explorability}
\label{par:explorability}

We demonstrate how the requirement \ref{req:Explorability}, explorability, is met by UE, using a dataset which allows for more direct interpretation of its density values than a seismic dataset: A scan of a Manganese nodule.
\footnote{\url{https://en.wikipedia.org/wiki/Manganese_nodule}} which can be found on the deep sea floor.

\begin{figure}
  \centering
  \begin{minipage}{0.49\textwidth}
    \subcaptionbox{ }{\includegraphics[width=\textwidth]{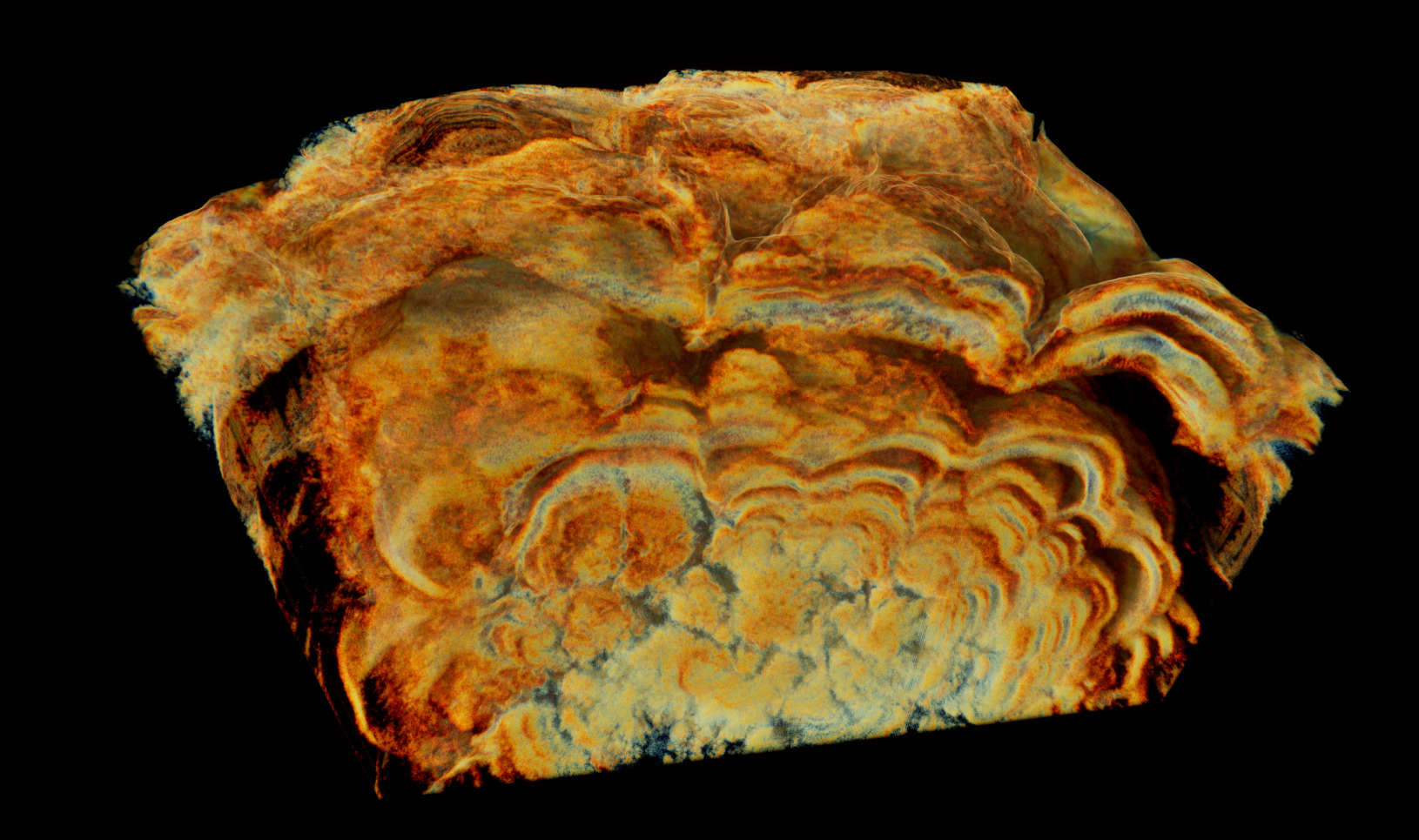}}
    \vspace{2mm} 
    \subcaptionbox{ }{\includegraphics[width=\textwidth]{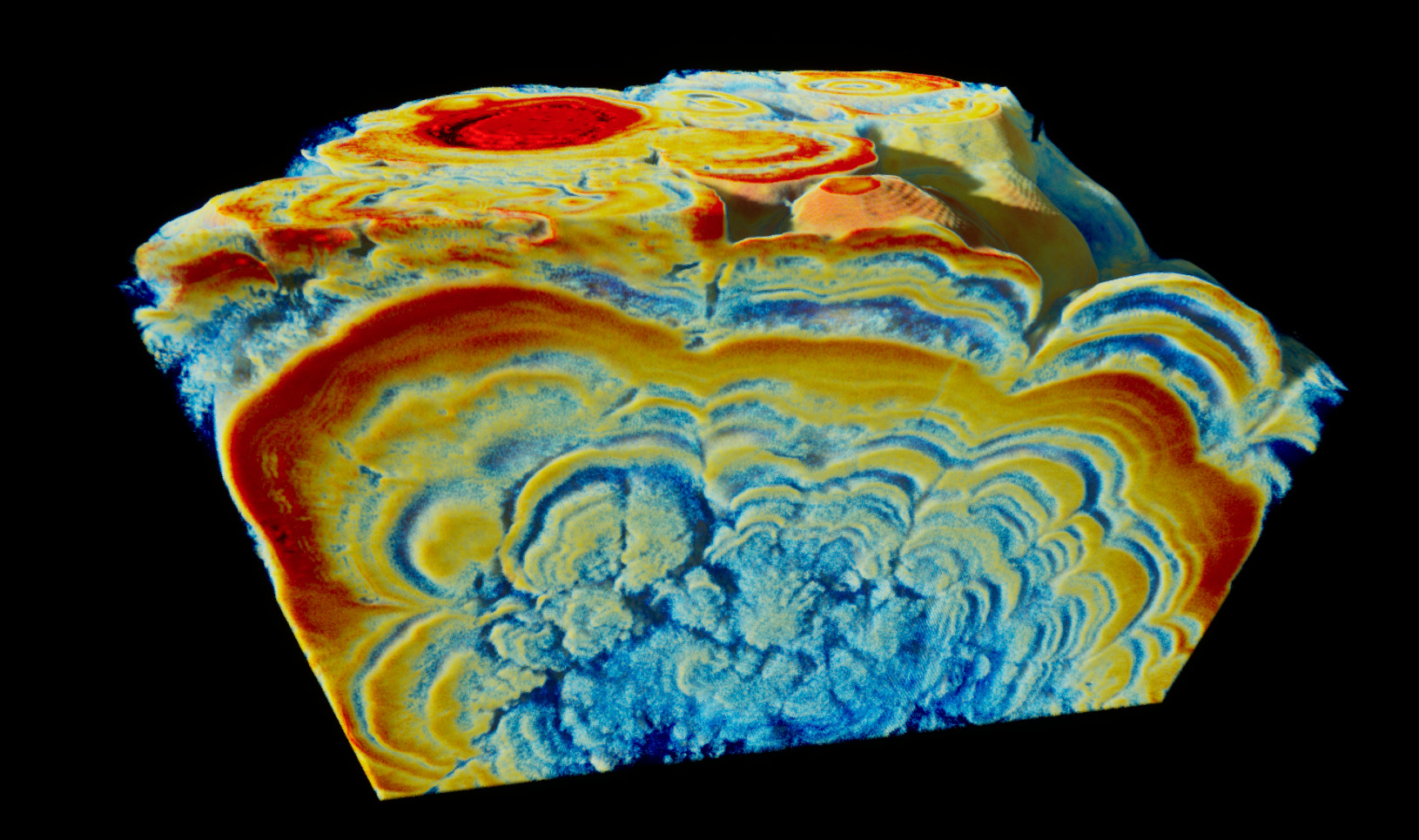}}
    \vspace{2mm}
    \subcaptionbox{ }{\includegraphics[width=\textwidth]{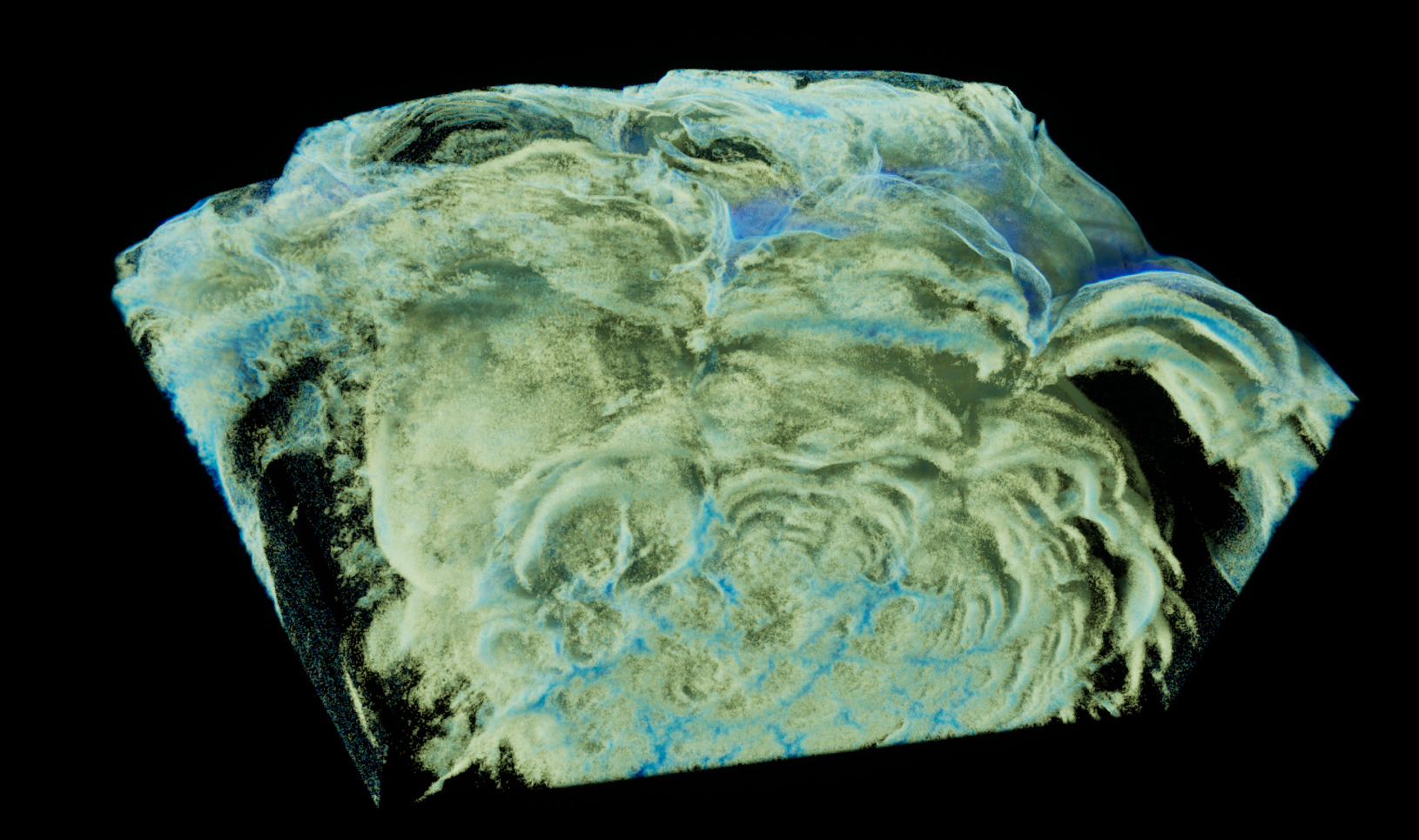}}
  \end{minipage}
  \hfill
  \begin{minipage}{0.49\textwidth}
    \subcaptionbox{ }{\includegraphics[width=\textwidth]{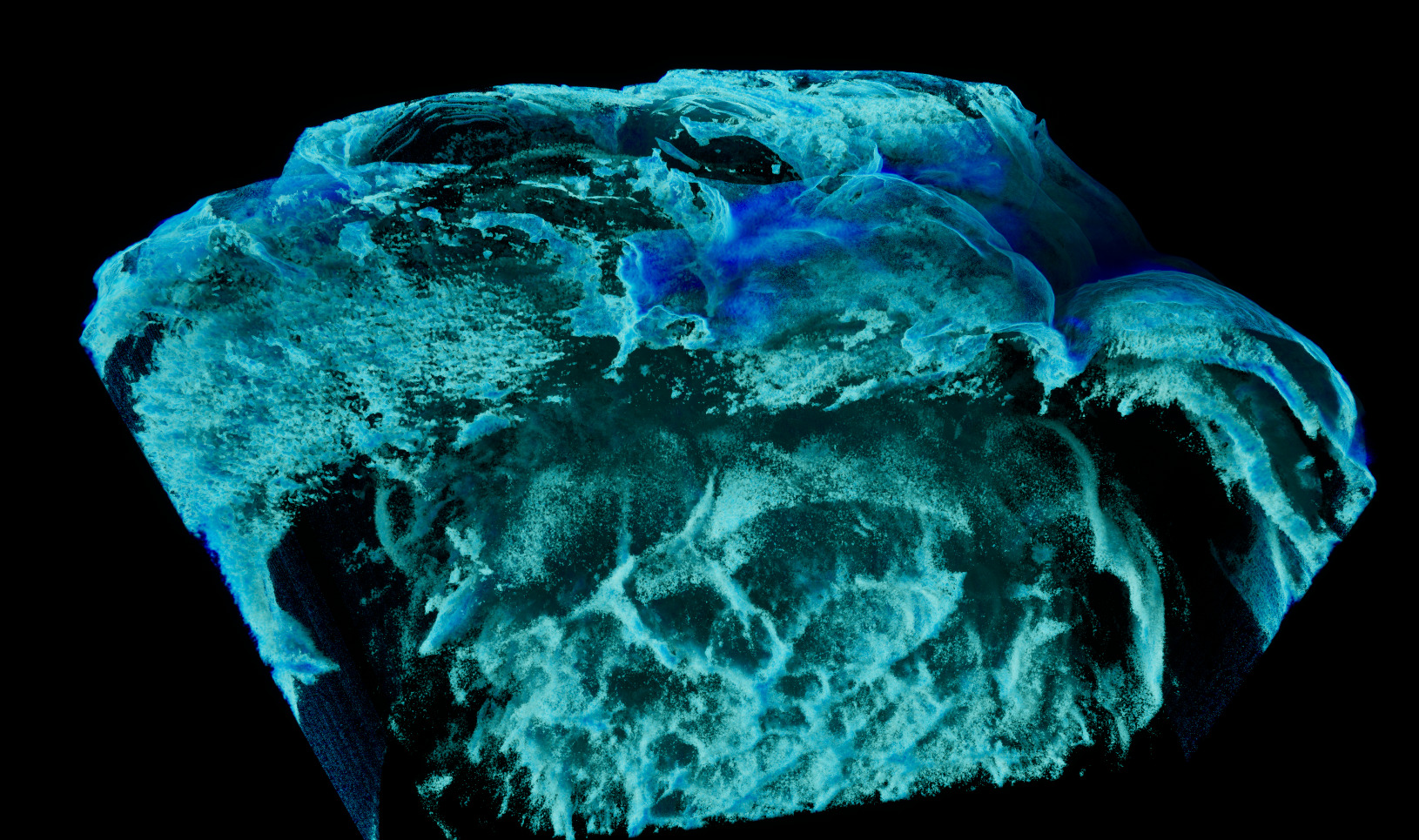}}
    \vspace{2mm} 
    \subcaptionbox{ }{\includegraphics[width=\textwidth]{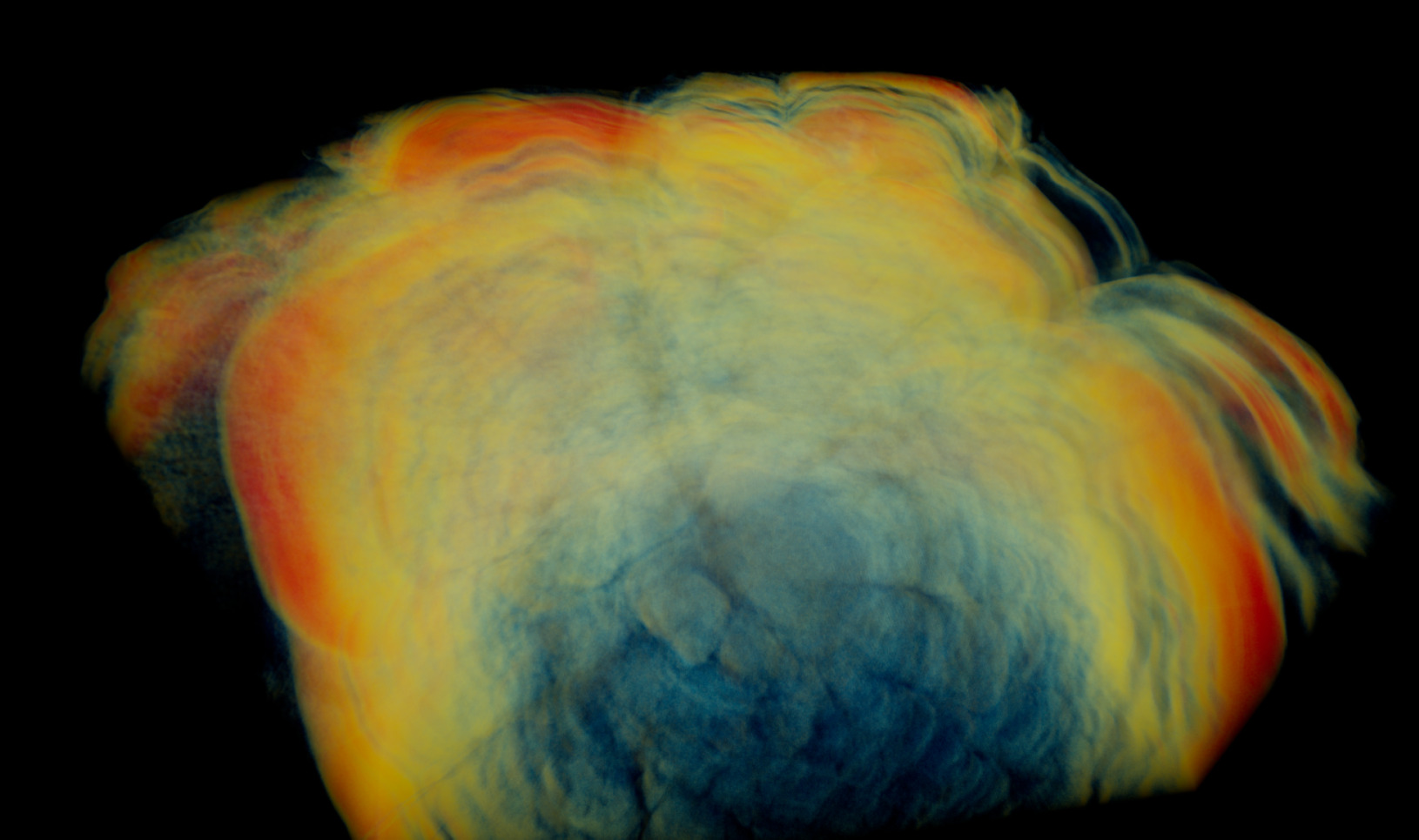}}
    \vspace{2mm}
    \subcaptionbox{ }{\includegraphics[width=\textwidth]{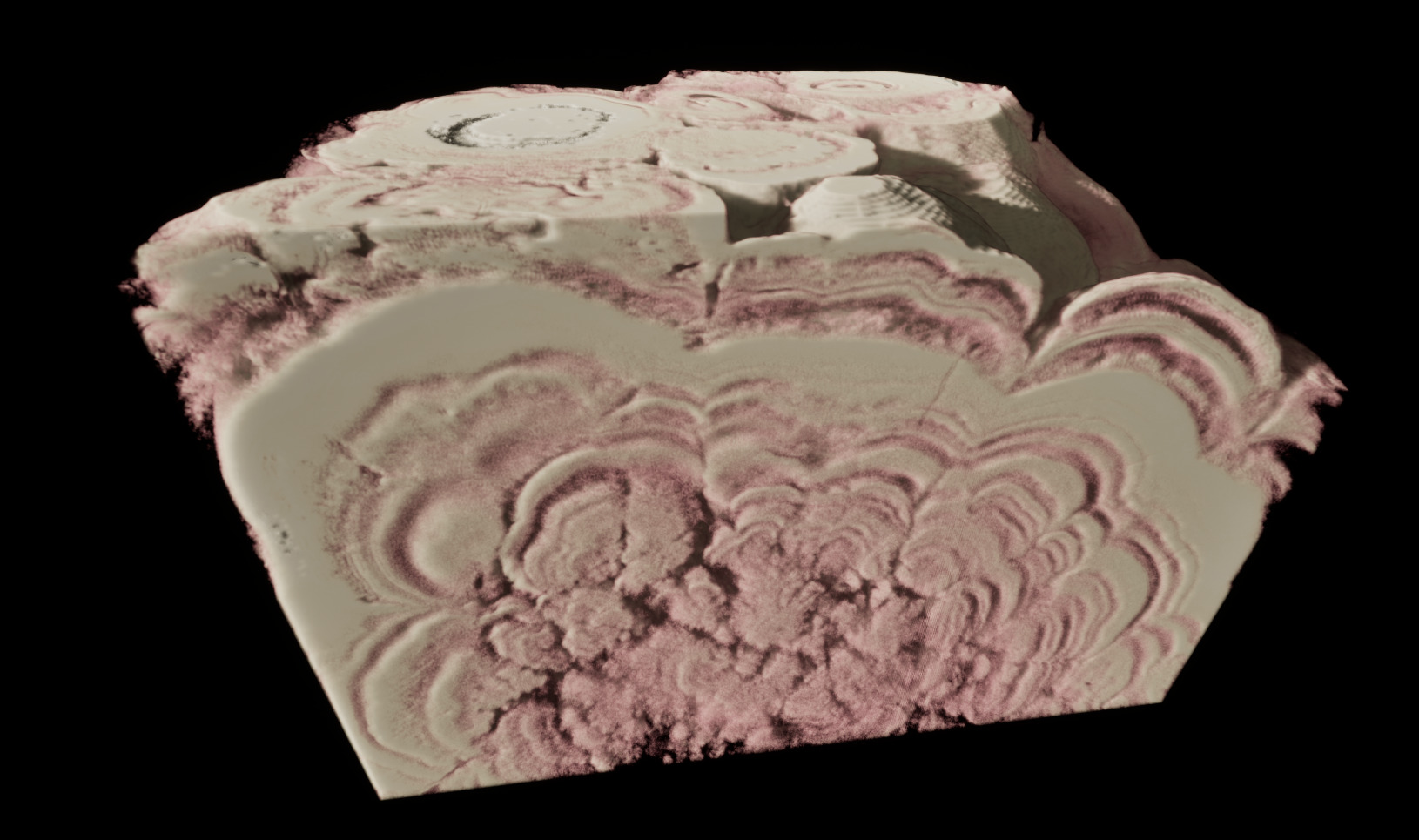}}
  \end{minipage}

  \caption{A scan of a manganese nodule using a Heterogeneous Volume, highlighting different features of the dataset by manipulating material parameters at runtime. The dataset has a resolution of
    $1536 \times 1536 \times 1034$ voxels and is discretized to 8 bit per voxel.
  }

  \label{fig:Manganese_nodules}
\end{figure}

Interactive exploration features like moving a cut plane or changing color maps (which can be used to encode transfer functions),
density \footnote{which influences transparency} and emission scales and ranges are easily implemented using the material system.
They are made available for dynamic changes at runtime (outside the editor mode) via
\lstinline|Material Parameter Collection|s (MPC) and \lstinline|Dynamic Material Instance|s.
Fig. \ref{fig:Manganese_nodules} demonstrates different possibilities to explore a dataset by
emphasizing different features of a dataset by changing material parameters at "game" runtime.
Additionaly, though only in the editor environment, UE's curve editor can be used to author a transfer function and see the update reflected to the rendering instantly. Unfortunately, graphical curve editing is not available at the "game" runtime.

\paragraph{Limits of Sparse Volume Textures}
\label{par:limits_SVT}

We would like to discuss the theoretical and practical limits of the SVT approach, as they enable an informed decision about future extension possibilities.
In particular, we elaborate on an third overflow issue we decided not to fix, as it would require some modification of the CPU-GPU upload code while extending the size limits of the SVT approach only in specific cases, or only slightly, before hitting another hard limit.
We argue that - using the above fixes - the SVT-approach uses the possibilities of DirectX 12 quite well, getting close to the API's theoretical limits.
Hence, we propose that the next step for loading large volume datasets into Unreal Engine is an extension to true out-of-core rendering, not only in the temporal, but also in the spatial dimensions.

Using DirectX 12 and reserved resources, it is possible to allocate volume textures with up to $2048^3=8$ gigavoxels, which fits our purposes better than the previous approaches, where we encountered practical limits at less than 1 gigavoxel. \newline
However, the \textit{ usable} payload data with the SVT approach are smaller:
Given that each $16^3$-voxel "sparse tile" needs to be padded to allow for correct trilinear interpolation at the tile borders, the usable payload in the texture is $2048^3 * 16^3/(1+16+1)^3 \approx 5.6$ gigavoxels. The tiles of the Mipmap hierarchy are also placed in the same texture, reducing the net payload data by another factor of $\sum_{n=0}^{\infty } (1/2^n)^3 \to 1.143$, leaving a net voxel count of about 4.9 G:
\begin{equation}
  2048^3 * 16^3/(1+16+1)^3 / 1.143 \approx 4.9 G
  \label{eq:netVoxelCount}
\end{equation}

We were surprised to see the engine crash for datasets approaching 2 gigavoxels, which led us to investigate and identify at least two places in the source code that caused an overflow at 2 gigabyte data size, due to \textit{signed} \lstinline|int32| usage. Our fixes allowed us to extend the importing capabilities towards the above theoretical limit.

The Kolumbo data set makes the engine allocate a volume texture with about $1872^3$ voxels for the tile data, reporting $6.7 \times 10^9 $ Bytes of reserved memory. So with this dataset, the SVT representation is \textit{not} really smaller than a hypothetical \textit{dense} texture would be
\footnote{
  $ 4211 \times 935 \times 1501 \times 1.143 \approx 6.7$ billion voxels $=$ 6.7 billion bytes for 8 bit voxel precision. It is a coincidence that the dense voxel representation has a similar size as its SVT representation: The "sparseness gain" of this compact dataset is "eaten up" by padding voxels and mipmap tiles.
}.
So in the case of the Kolumbo dataset, the advantage of the SVT is neither a performance gain nor less memory consumption, but circumventing the API limit of a maximum of 2048 voxels per dimension.

We encountered another crash due to an overflow using another almost fully dense dataset that has $2340 \times 2340  \times 177/(2^{30}) = 0.9$ gigavoxels.
When we tried to import it with 32 bit floating point precision, together with padding and Mipmaps, the \textbf{CPU}-side representation has size of
$0.9 GV \times 1.42 (\text{padding})\times 1.15 (\text{MIPs}) \times 4$ Bytes/voxel
$\approx$ 5.9 GB > 4 GB.
This is even caught in the code with the message \lstinline|"SVT streaming data overflowed the uint32 range!"|. Although it would be easy to extend the limit on the CPU-side, a buffer of this size must eventually be uploaded to the GPU.
On the GPU side, these buffers are accessed via \lstinline|ByteAddressBuffer|, i.e., for offsets > 4 GB, one would need to use 64-bit integer types. This will not only introduce a slowdown due to fewer 64-bit ALUs, but also the contents of the
\lstinline{TileDataOffsetsBuffer} would need an upgrade to offsets of 64-bit, introducing additional slowdowns. \newline
We did not analyze the CPU side of the upload code thoroughly. In principle, it should be possible to use a "sliding window" approach, mapping only a range within the CPU upload buffer to a smaller GPU upload buffer.
This would fit one aspect of the algorithm particularly well,
as according to the in-code comments, only $2^{27}$ elements can be uploaded at a single time, i.e. some kind of sliding window approach already exists. However, additional logic is needed to mark and reconstruct potentially overflown offsets.

We decided not to implement this extension yet, because for our use case, it would only slightly increase the practically usable voxel limit:
On the GPU side, the hard limit of the "payload" voxels is 4.9G (see equation \ref{eq:netVoxelCount}). \textit{This includes empty voxels in non-empty tiles}.
Assuming 4GB being the maximum size of the upload buffer: The maximum amount of non-empty voxels in the upload buffer is $4G/1.42/1.15= 2.5G$ (again due to padding and mip mapping). This implies that if the average occupancy of a tile is  $ < 2.5/4.9 \approx 50 \% $, the volume texture would be "blown up by empty waste voxels".
This means that expanding the limits of the upload buffer beyond 4GB will only benefit datasets whose non-empty tiles are relatively dense, while the sparser ones will hit the API limit due to empty voxels in the tiles.

The Kolumbo dataset is either completely empty (above the ground), or where it has data, it is relatively dense(average occupancy per tile is about 70\%). So it is of the kind of datasets that will profit from the extension, but only by a factor of $(2048/1871)^3 \approx 1.31$.

We conclude that extending the in-core implementation of SVT's by allowing the upload buffer to become bigger than 4GB  would mostly benefit datasets with high per-tile occupancy or high bit resolution, while there is little to no benefit for less compact datasets and those with lower bit resolution. These limitations can only be overcome by a true out-of-core implementation.

\section{Summary, Conclusion and Outlook}
\label{sec:conclusion}

We want to achieve an interactive volume rendering solution in our ARENA2 visualization dome
that supports huge datasets, fits well into the hard- and software infrastructure, and supports sustainable future extensions
for domain-specific processing, interaction, and visualization.

We chose the Unreal Engine as a basis for our endeavor, as it meets many of the requirements defined in Section \ref{sec:requirements}.

So far, we summarize our journey as follows.\newline
We started with the \textit{TBRayMarcher} (section \ref{subsec:tbrm}), which works relatively quickly out of the box with great performance, appealing visuals and some predefined transfer functions designed for a medical context. We encountered size limitations at about 1 gigavoxel. If we wanted to display larger datasets, we had to work around this limitation by chunking and displaying the chunks side-by-side, which still works with decent performance, but comes at the price of lighting artifacts at the chunk's borders. Furthermore, as the plugin is not maintained by Epic Games, users have to rely on updates by the author, which have happened in the past but are not guaranteed for the future.

We then tried using the Niagara Fluids plugin (Section \ref{subsec:niagara}) to display our data.
This worked -to a certain extent, and being part of the versatile Niagara system, additional processing and special visualization techniques might be added in the future by developing Niagara modules. The idea of authoring transfer functions using the Niagara interface was intriguing, but impractical for nontrivial datasets, not to mention that this feature is not available outside of the Unreal Editor. Unfortunately, the effective resolution per chunk (per Niagara system, in this case) needs to be quite low
\footnote{ $384^3$, about $4^3=64$ times smaller than the original dataset}
and even then, VRAM limits are hit quickly because of the severe simulation overhead.

We evaluated \lstinline|Sparse Volume Textures| and its \lstinline|Heterogeneous Volume| Actor, which - on paper - is a promising approach for our purposes. We were surprised to hit
a limit via a crash at about 2 gigavoxels of non-empty compressed voxel data, which is much lower than the theoretical limit of 4.9 gigavoxels (in form of $16^3$ tiles), see eq. \ref{eq:netVoxelCount}. We managed to identify the 2 gigavoxel limit as overflows of signed
\lstinline|int32| arithmetic ($2^{31}=2G$). We were then able to fix the overflow issues, import and render a 6 gigavoxel (GV) "dense" dataset,
whose internal representation turns out to be close to the 8GB texture limit for rendering ($6.7 \times 10^9$ byte), and very close to another potential overflow issue in the RAM-VRAM upload code, this time an \textit{unsigned}
$uint32$ overflow at $2^{32}=4$ GB.
We figured that it is not worth investigating to overcome a possible $uint32$ overflow, because it will extend the limits only by a small margin and for certain dense and concentrated voxel distributions.

In summary, we have found a solution to our initial problem of rendering large-scale volume data in our visualization dome that exploits the capabilities of current hardware and graphics APIs pretty well.

In the future, beyond implementing domain-specific processing and interaction with data in Unreal Engine, we plan to investigate \textit{sciview} for its multichannel and out-of-core rendering capabilities. Additionally, we aim to explore \textit{ParaView}'s potential to integrate its cluster mode for synchronized multichannel dome rendering with \textit{NVIDIA IndeX}'s cluster capabilities for distributed out-of-core rendering of large-scale volumetric data. Ultimately, we intend to explore extending Unreal Engine's volume rendering capabilities to support true out-of-core rendering.

\bibliographystyle{unsrt}
\bibliography{references}

\end{document}